\pdfoutput=1
\documentclass[useAMS,usenatbib]{mn2e}
\usepackage[pdftex]{graphicx}

\def\apj{ApJ}
\def\apjl{ApJL}
\def\aap{A\&A}
\def\mnras{MNRAS}
\def\nat{Nature}

\title[A new look at NICMOS transmission spectroscopy]{A new look at NICMOS transmission spectroscopy of HD189733, GJ-436 and XO-1: no conclusive evidence for molecular features}
\author[N. P. Gibson et al.]{
N. P. Gibson$^{1,2}$\thanks{E-mail: Neale.Gibson@astro.ox.ac.uk},
F. Pont$^{2}$ and
S. Aigrain$^{1,2}$
\\
$^{1}$Department of Physics, University of Oxford, Denys Wilkinson Building, Keble Road, Oxford OX1 3RH, UK\\
$^{2}$School of Physics, University of Exeter, Exeter, EX4 4QL, UK\\
}

\begin{document}
\topmargin -0.5in
\date{Submitted July 8th 2010}

\pagerange{\pageref{firstpage}--\pageref{lastpage}} \pubyear{2002}

\maketitle

\label{firstpage}

\begin{abstract}

We present a re-analysis of archival HST/NICMOS transmission spectroscopy of three exoplanet systems; HD 189733, GJ-436 and XO-1. Detections of several molecules, including H$_2$0, CH$_4$ and CO$_2$, have been claimed for HD 189733 and XO-1, but similarly sized features are attributed to systematic noise for GJ-436. The data consist of time-series grism spectra covering a planetary transit. After extracting light curves in independent wavelength channels, we use a linear decorrelation technique account for instrumental systematics (which is becoming standard in the field), and measure the planet-to-star radius ratio as a function of wavelength. We use a residual permutation algorithm to calculate the uncertainties, in an effort to evaluate the effects of systematic noise on the resulting transmission spectra. For HD 189733, the uncertainties in the transmission spectrum are significantly larger than those previously reported. We also find the transmission spectrum is considerably altered when using different out-of-transit orbits to remove the systematics, when some parameters are left out of the decorrelation procedure, or when we perform the decorrelation with quadratic functions rather than linear functions. Given that there is no physical reason to believe the baseline flux should be modelled as a linear function of any particular set of parameters, we interpret this as evidence that the linear decorrelation technique is not a robust method to remove systematic effects from the light curves for each wavelength channel. For XO-1, the parameters measured to decorrelate the light curves would require extrapolation to the in-transit orbit to remove the systematics, and we cannot reproduce the previously reported results. We conclude that the resulting NICMOS transmission spectra are too dependent on the method used to remove systematics to be considered robust detections of molecular species in planetary atmospheres, although the presence of these molecules is not ruled out.

\end{abstract}


\section{Introduction}

The number of known exoplanets is increasing rapidly, revealing a diverse range of systems with vastly different properties. Transiting exoplanets offer a unique opportunity to study the structure and composition of planets other than those in our own solar system, as transit light curves allow a measurement of the radius of the planet, and the complementary radial velocity technique measures the mass. From the derived density, the bulk composition of the planet may be inferred.

Transiting planets also provide the opportunity to measure the composition of planets' atmospheres. Typically, transit light curves are modelled assuming the planet as an opaque disk, whose size is defined by the altitude at which the atmosphere becomes opaque to starlight. However, the optical depth in the atmosphere is wavelength dependent, being sensitive to atomic and molecular absorption. Therefore, the transit depth and measured radius of a planet are wavelength dependent, and measuring the planetary radius as a function of wavelength may allow the detection of absorption features in the atmospheres, and thus enable observations to infer the presence of atomic and molecular species \citep[e.g.][]{Seager_2000,Brown_2001}. Hot Jupiters are a class of planets with masses similar to Jupiter and on very short-period orbits, therefore they are intensely irradiated resulting in very hot atmospheres. Consequently, their large atmospheric scale heights make them particularly good targets for this type of measurement.

There are two approaches to measuring the wavelength dependence of a planet's radius. Transmission spectroscopy consists of monitoring the transit with spectroscopic measurements, which can be subsequently split into separate light curves for each wavelength channel, allowing a measurement of transit depth at each wavelength. This has been used to detect various species in the atmospheres of hot-Jupiters, including HD 209458 \citep[e.g.][Na]{Charbonneau_2002} and HD 189733 \citep[e.g.][H$_2$O, CH$_4$]{Swain_2008}, and has also suggested a haze in the upper atmosphere of HD 189733 \citep{Pont_2008} at optical wavelengths. Another method to measure the wavelength dependence of a planet's radius, is to make multi-colour photometric observations of the transit light curve, and measure the transit depth for each. Spitzer observations of this type were used to infer the presence of H$_2$O in the atmosphere of HD 189733 \citep{Tinetti_2007}, although this is disputed by \citet{Ehrenreich_2007} and \citet{Desert_2009}.

These measurements have provided a wealth of information which feeds into theories of planetary atmospheres. However, both methods have advantages and disadvantages, and have been the subject of some controversy. For example, photometric measurements are much more straight forward, and the data reduction and analysis use established methods, and are therefore relatively robust against instrumental systematic effects. However, observations are (typically) taken during different transits, and stellar activity may cause variations of the measured planetary radius due to variable spots on the surface of the star, which change as the star rotates and reveals different hemispheres, and also evolve on timescales comparable to the period of typical transiting planets \citep{Mosser_2009}.

Transmission spectroscopy on the other hand, avoids this problem by simultaneously monitoring the transit light curve at different wavelengths. Whilst there is still a chromatic variance due to any spots on the stellar surface during a particular transit, this is a much smaller effect than if the surface spot distribution has changed. However, transmission spectroscopy is usually affected by systematic noise from the telescope and instrument \citep[e.g.][]{Pont_2008,Swain_2008}, often larger than the signal we are trying to measure. This is commonly treated by constructing a multi-linear model of the baseline function, which describes how the measured light curve of the star would behave in the absence of a planetary transit due to changes in the state of the optics and detector. This is built from a linear combination of parameters that describe the optical state of the system, such as the position of the spectral trace on the detector, the temperature of the detector, and in the case of Hubble Space Telescope (HST) observations, the orbital phase of the observatory. The baseline function is determined from the out-of-transit observations, and then used to decorrelate the light curve by projecting the function to the in-transit observations; however, the physical origins of these linear decorrelation models are poorly understood, and as we will see later, the choice of model has significant effects on the output transmission spectra.

HST transmission spectroscopy with NICMOS (Near Infrared Camera and Multi-Object Spectrometer) has led to some of the most detailed studies of exoplanet systems, but it in particular suffers from these systematics effects \citep[e.g.][]{Swain_2008,Pont_2009}, which are comparable to if not larger than the expected signal due to molecular absorption. Of the NICMOS observations currently in the literature, one group have claimed the detection of several molecules in the atmospheres of various hot Jupiters (e.g. \citealt{Swain_2008}, HD 189733; \citealt{Tinetti_2010}, XO-1). However, other analyses of NICMOS transmission spectroscopy, have been unable to untangle the instrumental systematic effects from wavelength dependent absorption from the planet \citep[e.g.][]{Pont_2009,Carter_2009}. Furthermore, for HD 189733, there have been significant disagreements with wavelength dependent photometric measurements of the light curve and transmission spectroscopy, with \citet{Sing_2009} failing to detect water in the atmosphere of HD 189733 previously reported by \citet{Swain_2008}, using narrow band photometric measurements with NICMOS. \citet{Tinetti_2010} attribute this to variable spot distributions for each of transit, arguing multi-colour photometric observations cannot reach the required accuracy for molecular spectroscopy, despite \citet{Sing_2009} accounting for this using continuous ground-based monitoring of the stellar flux.

As these results are vital to our understanding of planetary atmospheres, we must explore the possibility that the wavelength dependance of the planetary radius is no more than systematic noise in the detector. As the HST orbits with a period of $\sim$96 minutes, transit light curves are split into discrete orbits, typically sampling several out-of-transit portions of the light curve, used to establish the baseline function, and 1--2 in-transit portions. Trends are typically seen in each orbit, related to the orbital phase of the HST and its effects on the position of the spectral trace. Furthermore, as the optical system cannot be reset to exactly the same configuration for every orbit, systematic offsets between the flux levels in each orbit may also result. Efforts to gain a deeper understanding of NICMOS systematics and obtain photon limited photometry are ongoing \citep[e.g.][]{Burke_2010}. However, the offsets are not satisfactorily addressed in much of the literature, as any systematic offset may not be corrected for or indeed may be over-corrected for when the baseline function is projected to the in-transit portion of the light curve in the case of an imperfect baseline model. These in-transit offsets will be hidden from the light curve residuals when fitting the transit depth, and therefore not taken into account in the error analysis.

\begin{figure}
\includegraphics[width=84mm]{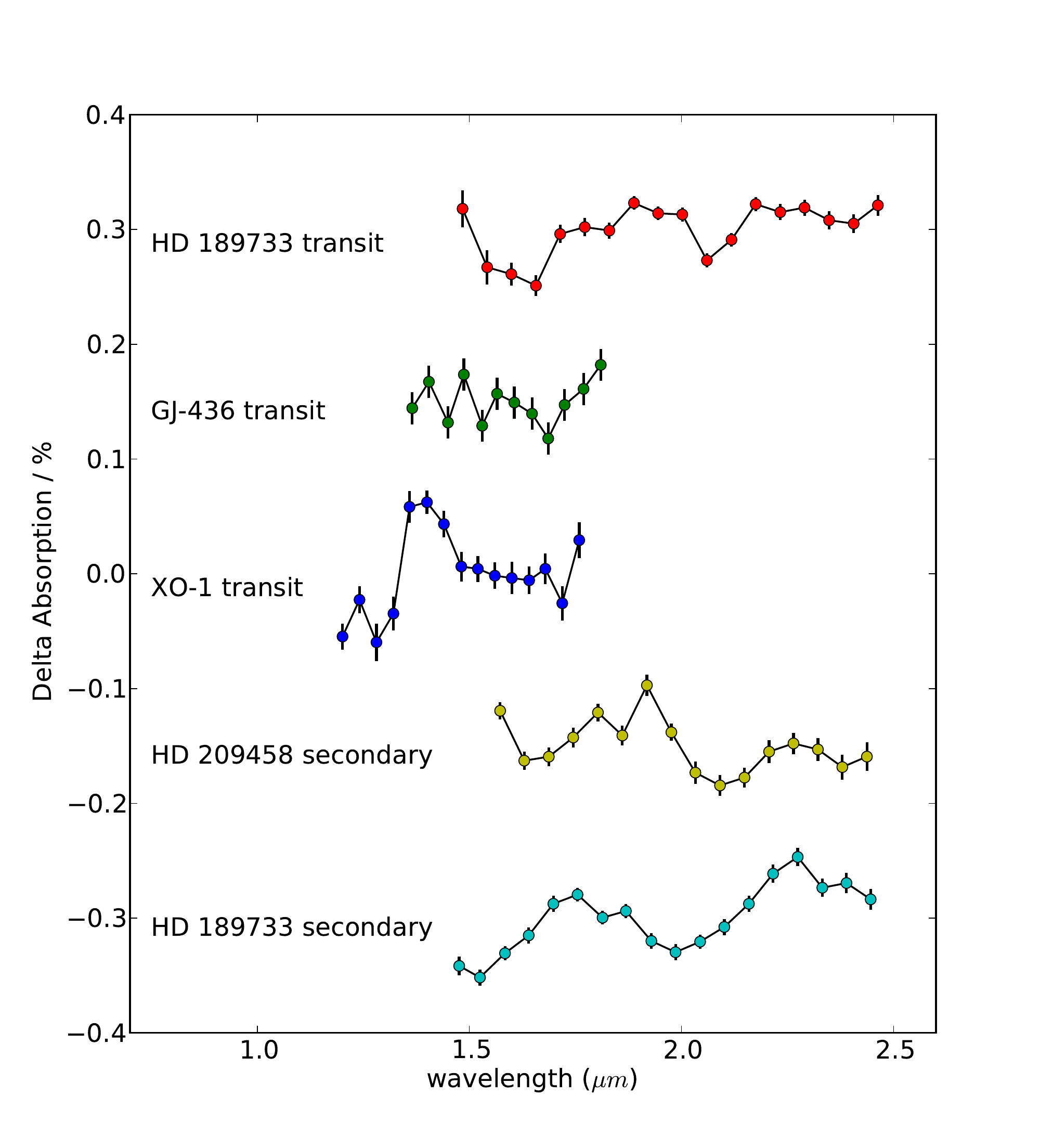}
\caption{Plot of some of the NICMOS transmission and emission spectra reported in the literature; transmission spectra for HD 189733 \citep{Swain_2008}, GJ-436 \citep{Pont_2009}, and XO-1 \citep{Tinetti_2010}, and dayside emission spectra for HD 189733 \citep{Swain_2009b} and HD 209458 \citep{Swain_2009a}. Each spectrum is plotted as \% absorption as a function of wavelength, but offset for clarity. The peak-to-peak amplitudes of all reported spectra are approximately equal, and also show similar features. }
\label{fig:NICMOS_data}
\end{figure}

Fig.~\ref{fig:NICMOS_data} shows a plot of the NICMOS spectra of transiting planets reported in the literature to date\footnote{For XO-1 and HD 209458 secondary, only every 5th point is plotted for comparison, because a boxcar smoothing was applied to the published data.}; transmission spectra for HD 189733 \citep{Swain_2008}, GJ-436 \citep{Pont_2009}, and XO-1 \citep{Tinetti_2010}, and dayside emission spectra for HD 189733 \citep{Swain_2009b} and HD 209458 \citep{Swain_2009a}. The emission spectra are obtained in much the same way as the transmission spectra, by observing an eclipse of the planet, and using the same linear decorrelation methods to correct for the baseline function. It is clear that all five spectra show approximately the same peak-to-peak amplitude, and have wavelength dependance on similar scales. This is perhaps surprising, given the planets have vastly different masses, radii and irradiation levels. Transmission and emission spectra also show the same amplitudes. Whilst this is by no means evidence that the spectra are the result of systematic noise in the detector, and the similarity could be coincidental, it is nonetheless wise to explore this possibility. It is therefore important that the results reported from NICMOS transmission spectroscopy be re-analysed to confirm previous detections of molecules, in particular because the results are dependent on a complex decorrelation procedure, with an arguably \emph{ad hoc} model.

Here, we present a re-analysis of the three NICMOS transmission spectra shown in Fig.~\ref{fig:NICMOS_data}. Sects.~\ref{sect:HD189733}, \ref{sect:GJ436} and \ref{sect:XO1} describe the data reduction, analyses and results of the HD 189733, GJ-436, and XO-1 data, respectively. Finally, Sect.~\ref{sect:summary} summarises and presents our conclusions.

\section{HD 189733}
\label{sect:HD189733}

\subsection[]{Observations}
\label{sect:HD189733_observations}

A transit of HD 189733 was monitored on 25 May 2007 with HST/NICMOS, using the G206 grism covering the wavelength range 1.4 -- 2.5 \micron. Analyses of these data were first reported in \citet[][hereafter S08]{Swain_2008}. As HD 189733 is not in the continuous viewing zone of the HST, the transit was observed over five half-orbits ($\sim48$ minutes each), consisting of 638 spectra in total, all with exposure times of 1.624 seconds. The first, second, fourth and fifth orbits cover the out-of-transit part of the light curve and consist of 119, 128, 131 and 130 observations, respectively. These data are required to determine the photometric baseline. Only the third orbit was taken in-transit, and consists of the remaining 130 images. In addition to the spectra, some exposures were taken at the beginning of the first orbit to enable wavelength calibration.

\begin{figure}
\includegraphics[width=84mm]{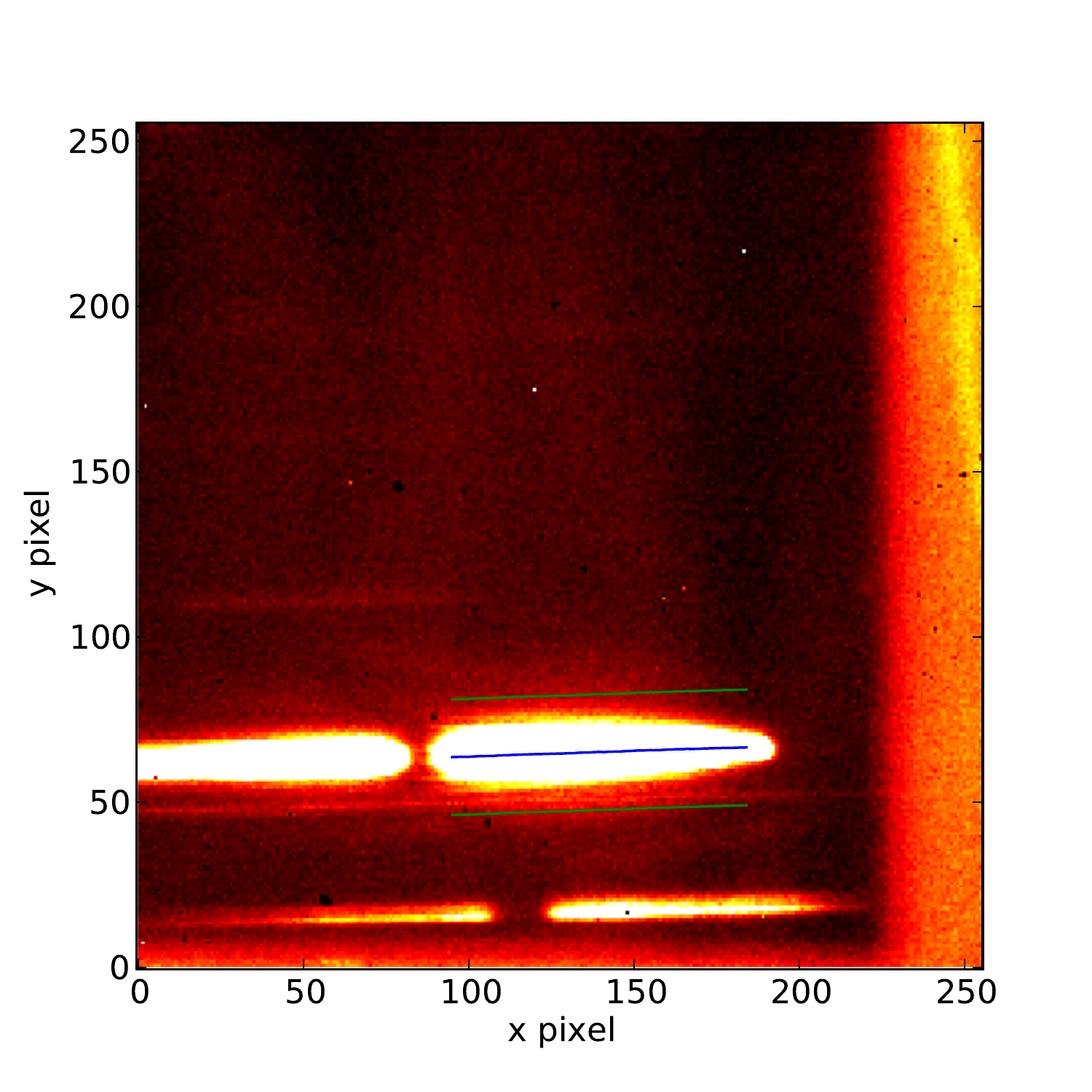}
\caption{NICMOS image of HD 189733 taken with the G206 grism. The blue line marks the centroids of the first-order spectrum of HD 189733, with the green lines flanking representing the extent of the extraction region. Immediately left is the second-order spectrum, and below is a companion star to HD 189733. The bright band at the right of the detector is a feature of G206 grism spectra, caused by the warm edge of the aperture mask.}
\label{fig:HD189733_image}
\end{figure}

The calibrated images, one of which is displayed in Fig.~\ref{fig:HD189733_image}, include all basic calibrations except for flat-fielding. Correctly flat fielding spectroscopic data requires taking into account the wavelength dispersion and position of the source, and therefore a different flat-field correction would be required at each point along the spectral trace of the target. S08 argue that flat-fielding is not required for differential spectroscopy, as each light curve will be normalised before measurement of parameters from the light curves. However, strictly speaking flat-fielding is required to accurately determine the background value in each wavelength channel. This is particularly important, as underestimating or overestimating the background results in a variable transit depth, which is exactly what we are trying to measure. Furthermore, the G206 grism has a relatively high background compared with NICMOS's other two grisms (about $800$ electrons per pixel per exposure), due to thermal background emission from HST. Whilst the background is very stable temporally, unfortunately, it does significantly vary spatially over the detector by as much as 200 electrons per pixel per exposure. This will particularly effect the light curves extracted from the edge of the spectrum, where the overall counts are much lower. We did try flat-fielding each image using a flat-field taken with the G206 grism in place. This corrects for diffuse light over the whole detector. This will not effect the depth of the final normalised light curves except through the background value. This process does not provide a satisfactory background correction, and the analyses described below were attempted with and without flat-field corrections. We also note that the background varies quite smoothly, and therefore is unlikely to be responsible for any narrow `features' seen in the transmission spectrum, but can certainly affect the overall depth and shape.

Prior to extracting the spectra, we identified the defective pixels flagged by the calibration pipeline, and additional significant outliers, and corrected for them by replacing the pixel value with an interpolation of the surrounding eight pixels. A relatively small number of pixels required corrections, and in fact this process has very little effect on the output spectra. A 1D spectrum was then extracted as follows for each image. For each column along the spectral trace in the dispersion axis (x), a centroid along the spatial axis (y) was calculated to determine the position of the spectrum. A sum of 35 pixels along each column centred on this position was used to extract the flux for each wavelength channel, after subtraction of a background value for each pixel. A width of 35 pixels was chosen to minimise the RMS in the white light curve of orbits 2, 4 and 5. This was repeated for 90 pixel columns along the dispersion axis. The extraction regions used are marked in Fig.~\ref{fig:HD189733_image}. For the background subtraction, we experimented with various techniques. First, a global background subtraction was used, similarly to S08. The background was taken to be the average of a large un-illuminated region above the spectral trace (we tested using different regions). This is not ideal, as previously mentioned the background varies spatially over the detector. As a first order correction, we instead calculated the background separately for each pixel column, as the average value of the un-illuminated region above the spectral trace along the column. Again we note that this does not provide a satisfactory correction, as the background varies along both x and y. Both global and wavelength dependent corrections were used for the subsequent analysis. For the remainder of this paper, we will display results from data that is not flat-fielded, and using separate columns for background subtraction, but our conclusions remain the same for each case.

\begin{figure}
\includegraphics[width=84mm]{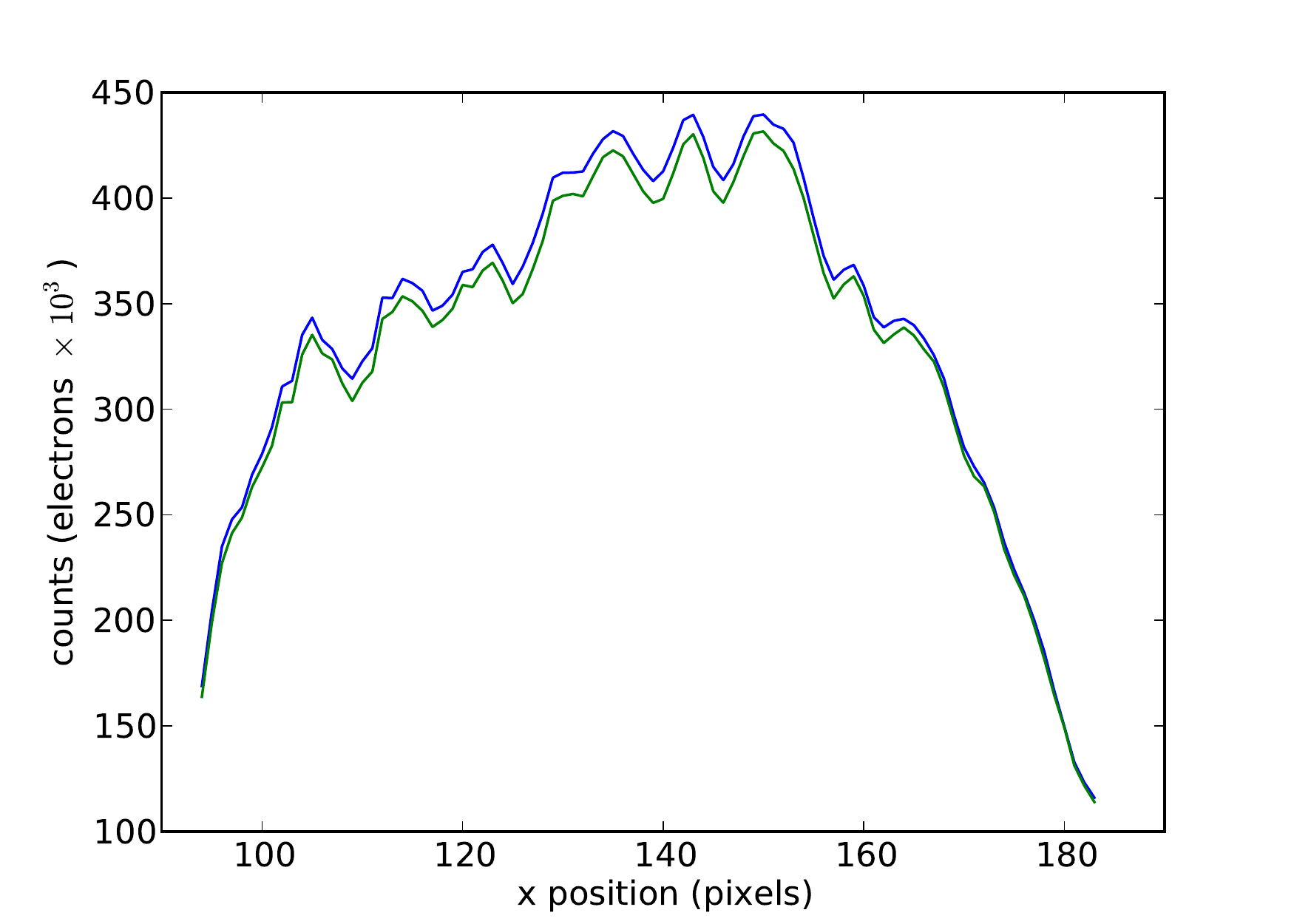}
\caption{Extracted 1D spectra of a typical in-transit (green) and out-of-transit (blue) observation of HD 189733 showing the number of electrons collected per pixel channel. The wavelength decreases with increasing x position.}
\label{fig:HD189733_1D_spectra}
\end{figure}

\begin{figure}
\includegraphics[width=84mm]{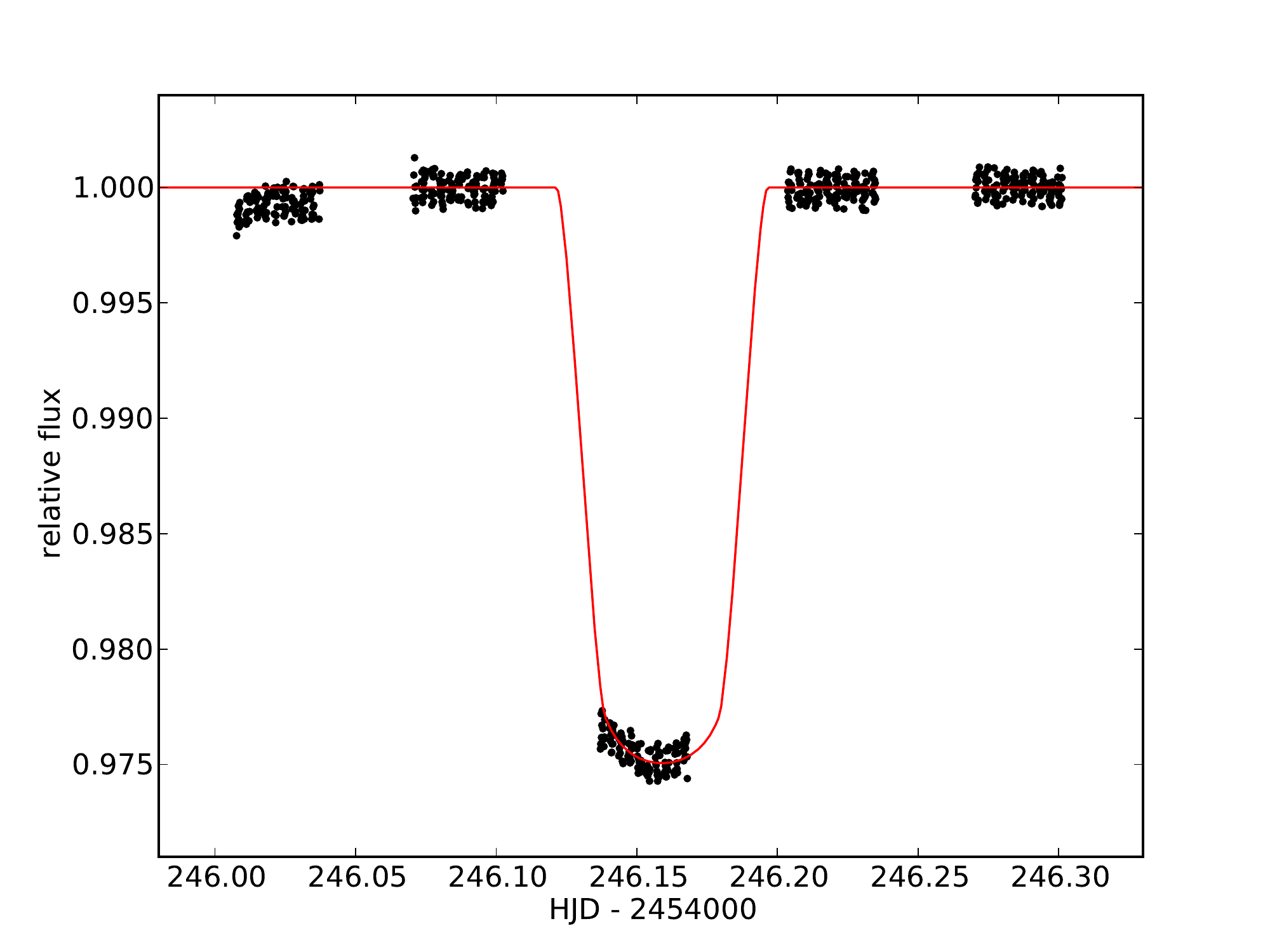}
\caption{Raw `white' light curve of HD 189733 obtained by integrating the flux from each spectra over all wavelengths, showing the sampling of the transit, alongside its best-fit model. While all orbits are seen to suffer from systematics, orbit one exhibits by far the largest, commonly attributed to spacecraft `settling', and is excluded from subsequent analysis.}
\label{fig:HD189733_wlc}
\end{figure}

Extracted spectra from a typical in-transit and out-of-transit observation are shown in Fig.~\ref{fig:HD189733_1D_spectra}, giving approximately $430\,000$ electrons in the brightest pixel channel, and approximately $120\,000$ electrons in the faintest channel. The features in the spectra do not correspond to real  stellar features, but result in variation of the sensitivity of the grism with wavelength. Each wavelength channel in the 1D spectra is then used to construct a time-series, after binning in 5 pixels along the dispersion direction, resulting in 18 light curves. A `white' light curve was constructed by integrating the flux over the entire wavelength range for each image. This was extended to include 110 pixel columns, which minimises systematics arising from small changes in the position of the spectral trace. This is plotted in Fig.~\ref{fig:HD189733_wlc}, and shows the sampling of the light curve over the five orbits. Systematics are evident in each orbit, but particularly for orbit one. This is commonly found in similar NICMOS data (see e.g. Sects.~\ref{sect:GJ436_obs} and \ref{sect:XO1_obs}). It is attributed to spacecraft `settling' \citep[e.g. S08;][]{Pont_2009}, and orbit one is excluded for the remainder of this work.

\begin{figure}
\includegraphics[width=84mm]{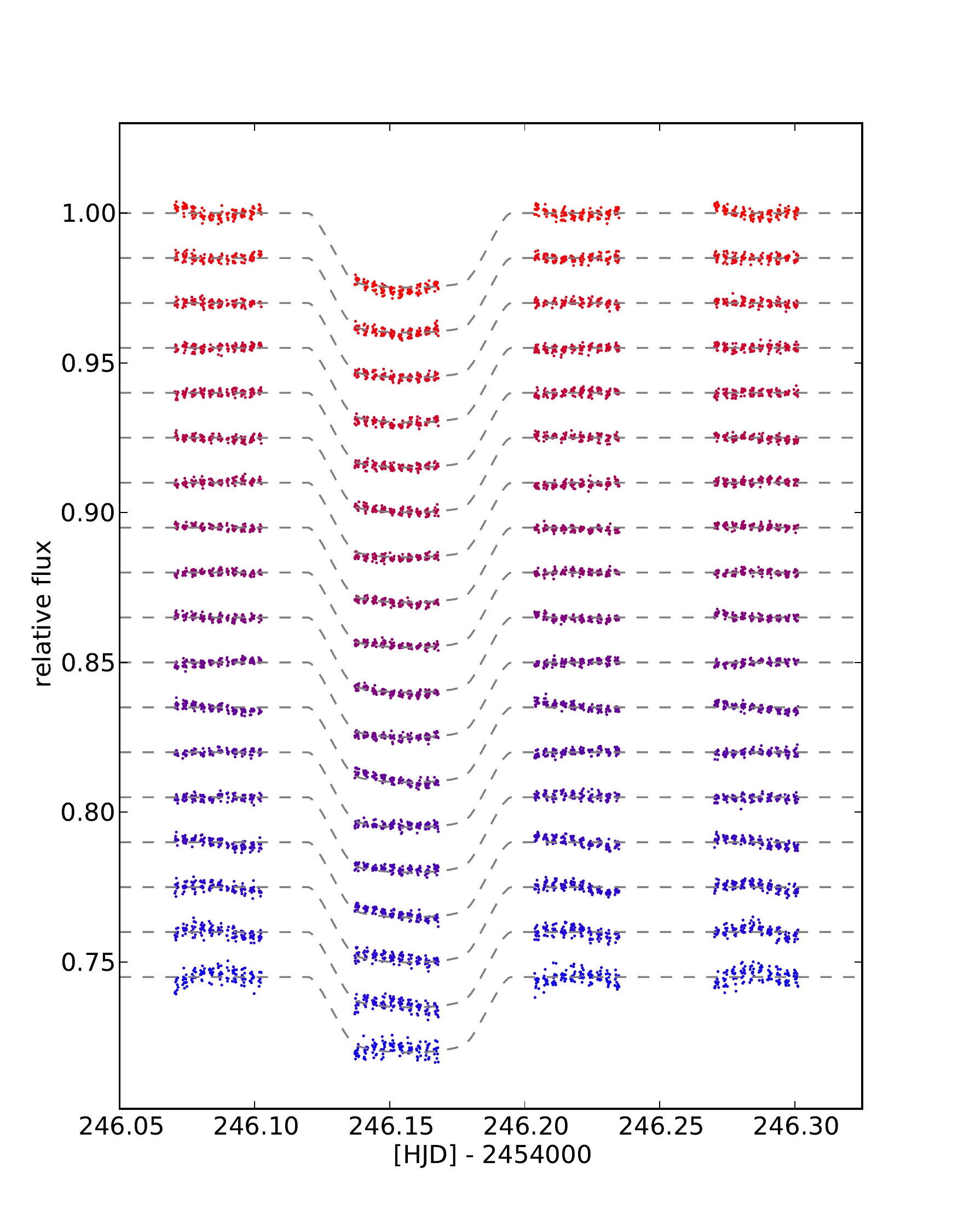}
\caption{Raw light curves of HD 189733, for each of the 18 wavelength channels from 2.50\,\micron~(top) to 1.48\,\micron~(bottom), after normalising by fitting a linear function through orbits 2, 4 and 5. Orbit 1 is not plotted as it is discarded for the analysis. Clearly, time-correlated noise is present in each light curve, and must be removed to measure the transmission spectrum. The dashed grey lines show a transit model generated for HD 189733, used to guide the eye.}
\label{fig:HD189733_normalised_lightcurves}
\end{figure}

The raw light curves are shown in Fig.~\ref{fig:HD189733_normalised_lightcurves} for each of the 18 wavelength channels, after normalising by fitting a line through orbits 2, 4 and 5, and dividing the light curves by this function. The light curves are clearly seen to exhibit strong time-correlated noise, which needs to be removed if we are to measure the transmission spectrum at the level required to detect molecular features. The systematics may be understood to arise from small motions of the spectra across the detector, related to the orbital motion of the HST. Referring back to Fig.~\ref{fig:HD189733_1D_spectra}, the wavelike features in the spectra will move into different wavelength channels and cause features on short wavelength intervals, even if the stellar spectrum and flat-fielding are smooth.

\begin{figure}
\includegraphics[width=84mm]{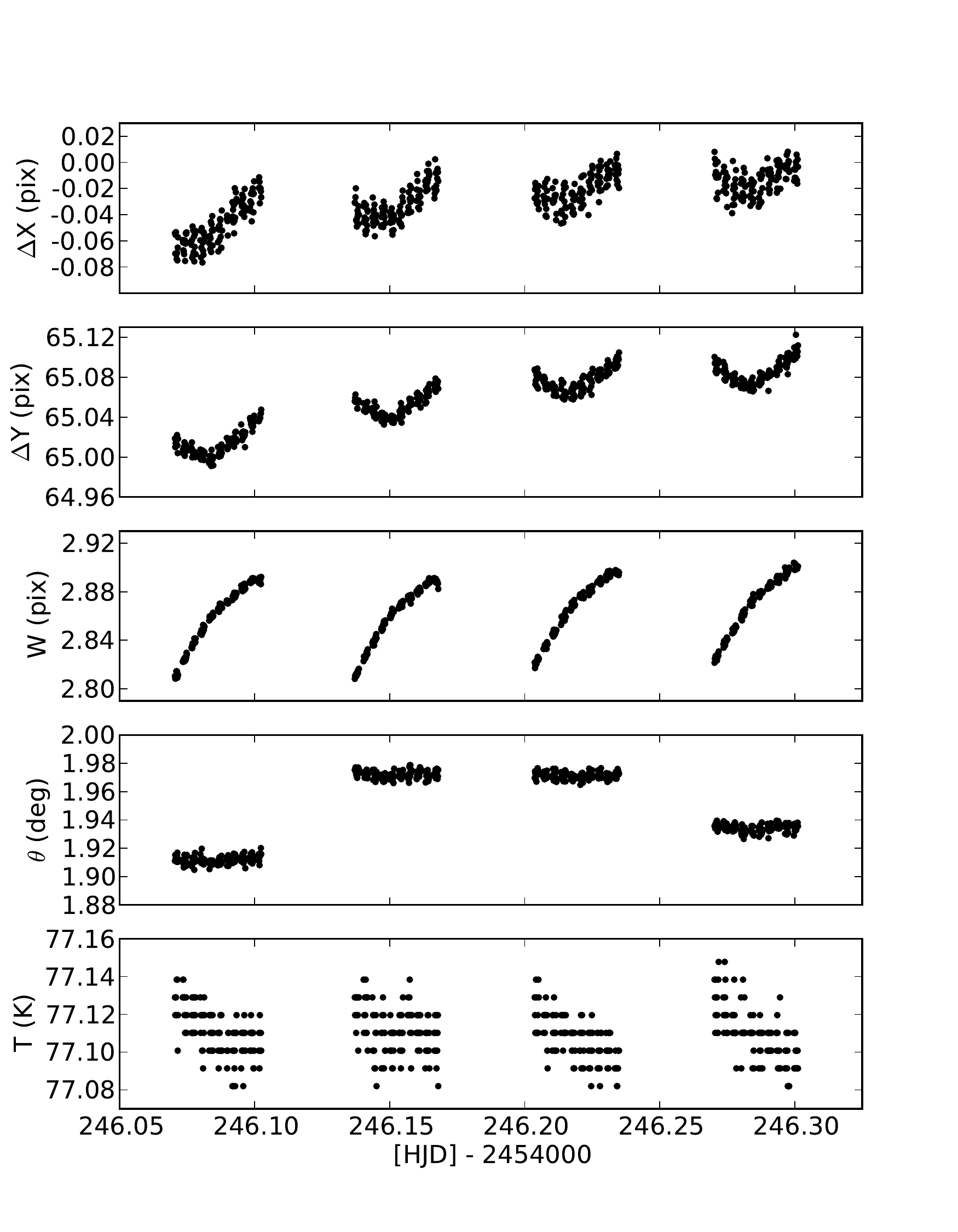}
\caption{Extracted state parameters for HD 189733 observations for orbits 2--5. For orbit 3 we reconstruct the model baseline as a linear function of all state parameters, plus a constant flux level, and the orbital phase of the HST and its square. The coefficients for each parameter of the model are determined from orbits 2, 4 and 5.}
\label{fig:HD189733_decorr_parameters}
\end{figure}

The optical state parameters, described by S08, were therefore measured in an effort to model and remove these systematic effects. We extracted the shift of the spectral trace along the x axis ($\Delta X$) by cross-correlation of the 1D spectra, the shift of the position of the y axis by averaging the centroids determined earlier ($\Delta Y$), and the angle the spectral trace makes with respect to the x-axis ($\theta$) by fitting a line to the centroids of the spectra. The width ($W$) of each spectrum was also measured by fitting gaussian functions along each extraction column, and taking the average. In the absence of a direct measurement of the  temperature ($T$) of the detector, or a proxy calculated from the bias levels \citep{Gilliland_2003}, a proxy for this was taken as the temperature of the NIC1 mounting cup (S08). However, this is not monitored at a high enough precision to accurately monitor the temperature. The temperature is an important parameter to describe the state of a NIR detector, and this is likely an important limitation of NICMOS analysis. A plot of the optical state parameters and the detector temperature is shown in Fig.~\ref{fig:HD189733_decorr_parameters}.

It is important to note that we extract our parameters slightly differently from S08 in some cases. In particular we used an averaged value for the shifts in position $\Delta X$ and $\Delta Y$ per image, in order to monitor the overall optical state of the detector, and therefore had identical parameters for all wavelength channels, unlike S08. S08 determined x and y positions by fitting spectra along diagonals and rotating back along the two nominal axis, resulting in different parameters for each wavelength channel, and in some cases different trends. In either case, the parameters show a similar dispersion and amplitude, and should allow a similar removal of the systematics.

\subsection[]{Analysis}
\label{sect:HD189733_analysis}

Each light curve was decorrelated using the multi-linear baseline model from S08. This assumes that the baseline function can be constructed by a linear combination of the optical state vectors $\Delta X$, $\Delta Y$, $W$ and $\theta$, the proxy for the detector temperature $T$, and also the orbital phase of the HST ($\phi_H$) and its square ($\phi_H^2$). In addition, we used a constant vector to represent the out-of-transit flux level ($f_{0}$), as clearly the overall flux level is not given by a linear function of the state variables, but rather the state variables are used to correct for variations from the expected out-of-transit flux. These parameters are hereafter collectively referred to as the `decorrelation parameters'. For each wavelength channel, the out-of-transit baseline flux is given by
\[
y_i = \sum_{k=1}^N \beta_k X_{i,k} + \epsilon_i
\]
at time $i$, where $\beta_k$ are the coefficients for each of the $N$ decorrelation parameters (in this case; $f_0$, $\Delta X$, $\Delta Y$, $W$, $\theta$, $T$, $\phi_H$ and $\phi_H^2$), $X_{i,k}$ is the value of each decorrelation parameter $k$ for time $i$, and $\epsilon_i$ the corresponding residual. This may be easily written in matrix form as
\[
\vec y = \mathbf{X} \vec\beta + \vec\epsilon,
\]
where $\mathbf{X}$ is the state matrix containing the measured de-correlation parameters, $\vec\beta$ is a vector containing the coefficients of each decorrelation parameter, and $\vec\epsilon$ are the residuals. The best-fit coefficients $\hat\beta$ are then found by linear least squares
\[
\hat\beta = (\mathbf{X}^T\mathbf{X})^{-1}\mathbf{X}^T \vec y,
\]
for orbits 2, 4 and 5 \emph{only}. The model baseline function $\vec b$ can then be reconstructed for \emph{all} orbits from a linear combination of $\hat\beta$ and $\mathbf{X}$;
\[
\vec b = \mathbf{X}\hat\beta.
\]
Each light curve is then decorrelated by dividing through by the model baseline function. This technique can only be expected to work if we interpolate for the in-transit orbit, and if the baseline function is well represented by the linear model over the range of the decorrelation parameters.

\begin{figure}
\includegraphics[width=84mm]{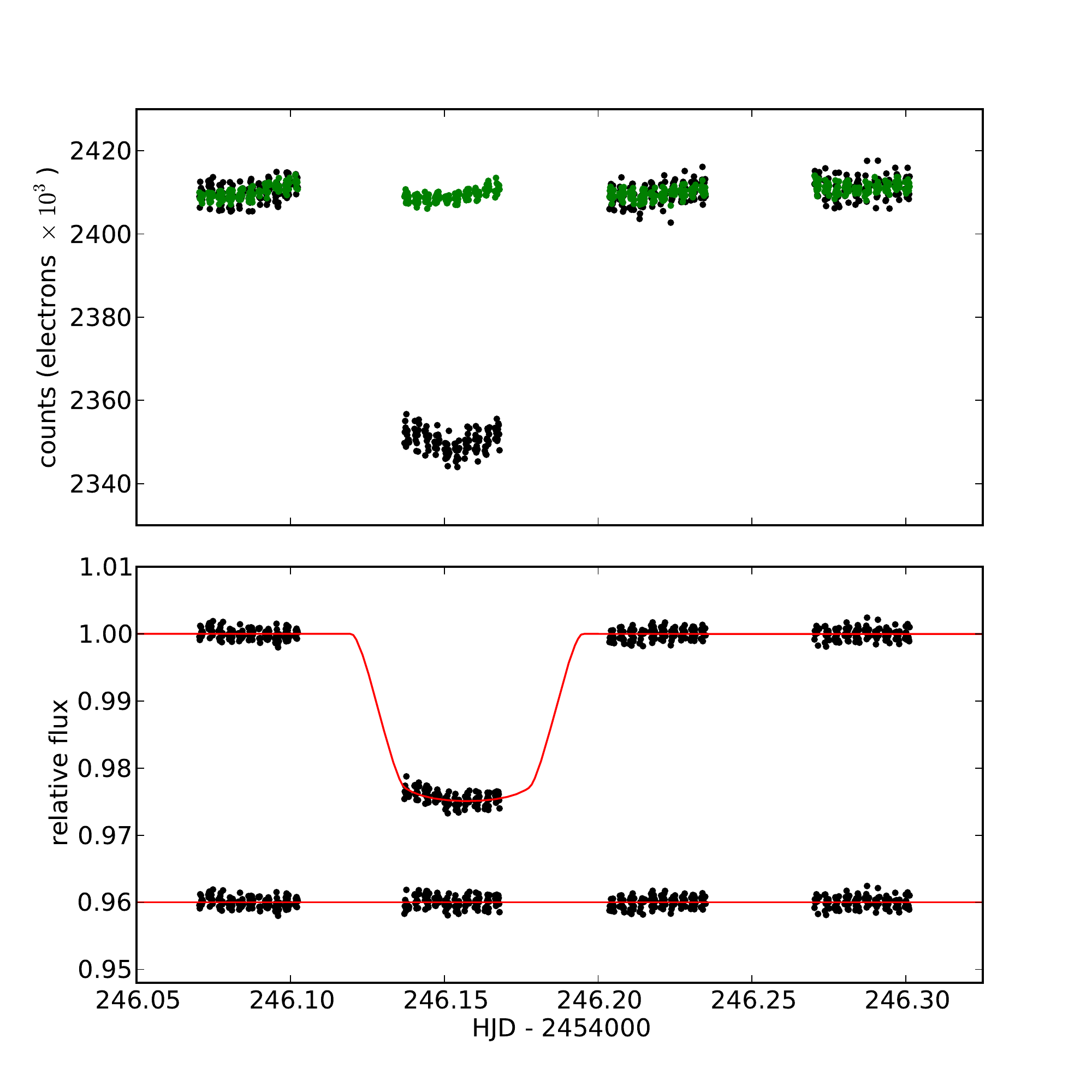}
\caption{Example of the decorrelation procedure on one of the wavelength channels. The top plot is of the raw light curve in electron counts. The green points represent the reconstructed baseline function ($\vec b$) used to decorrelate the light curve. The bottom plot shows the decorrelated light curve along with its best-fit transit model, used to determine the planet-to-star radius ratio. Whilst the out-of-transit orbits are usually whitened, residual systematic noise is often seen in the in-transit orbit, showing the linear baseline model is not robust.}
\label{fig:HD189733_decorr_eg}
\end{figure}

\begin{figure}
\includegraphics[width=84mm]{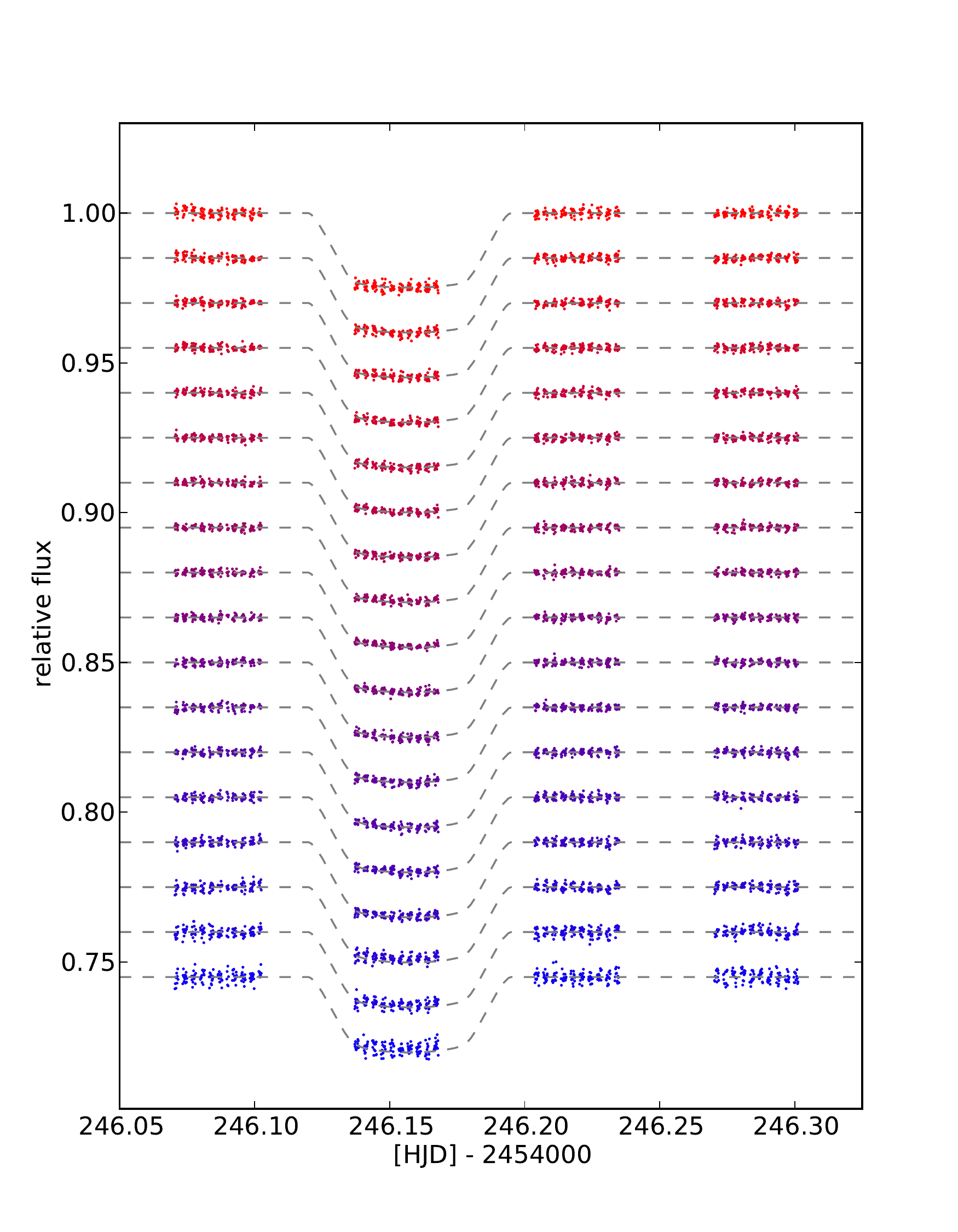}
\caption{Decorrelated light curves for each of the 18 wavelength channels. Much of the time-correlated systematic noise is removed from the light curves, as seen from a direct comparison with Fig.~\ref{fig:HD189733_normalised_lightcurves}. However, some of the correlated noise remains particularly in the in-transit orbit, and must be taken into account in the error analysis of the transmission spectrum. The dashed grey lines again show a transit model generated for HD 189733.}
\label{fig:HD189733_decorr_lightcurves}
\end{figure}

For this analysis, each light curve is treated \emph{independently}, with $\hat\beta$ calculated separately for each. An example of this decorrelation process on one of the wavelength channels is shown in Fig.~\ref{fig:HD189733_decorr_eg}, and all of the decorrelated light curves are shown in Fig.~\ref{fig:HD189733_decorr_lightcurves}. S08, who use a similar procedure to this, conclude that it satisfactorily removes time-correlated systematics from the light curves. Whilst our results agree with this conclusion for the out-of-transit orbits, residual systematics are clearly visible in the in-transit orbit, which is the important one for determining the radius ratio.

In order to obtain the transmission spectrum from the light curves, a transit model must be fitted to each decorrelated light curve to measure the planetary radius. We used the transit model described in \citet{Gibson_2008}, which assumes a circular orbit to calculate the normalised separation of the star and planet centres ($z$) as a function of time, using the orbital parameters and the masses and radii of the star and planet. The analytic models of \citet{Mandel_Agol_2002} were then used to calculate the light curves from $z$, the ratio of the planet-to-star radii $\rho$, and the limb darkening parameters. The limb darkening parameters were calculated in each of the wavelength channels for a quadratic limb darkening law (D. Sing, private communication) using the methods described in \citet{Sing_2010}. The orbital parameters of the system and the stellar mass and radius were held fixed at the values given by \citet{Pont_2008}, and the central transit time was determined from the ephemeris. Each decorrelated light curve was then fitted for $\rho$, to obtain the transmission spectrum.

To calculate the uncertainty in $\rho$ for each wavelength channel, we used a residual permutation (or `prayer bead') algorithm \citep[see e.g.][]{Gillon_2007,Southworth_2008}, similar to that used in \citet{Gibson_2009,Gibson_2010}. This method accounts for the correlated noise in the light curve by reconstructing light curves by combining the best-fit model and its residuals, but shifting the residuals before combining them to determine the effects of the correlated noise on the determined parameters. This method preserves both the correlated and random noise in the resampled light curves, and therefore both are taken into account when determining uncertainties.

The residual permutation was applied to the raw light curves prior to performing the decorrelation procedure, in order to take into account uncertainties from both the linear decorrelation and the light curve fitting. The best fit baseline function $\vec b$ and transit model $\vec m_{bf}$ were determined as before, and the residuals from the fit $\vec r$ are used to reconstruct the light curve, but each time the residuals are shifted by a random offset to give $\vec r_p$. Any residuals that fall off the `edge' are looped back to the beginning. The new light curve $\vec y_p$ is then reconstructed by adding the shifted residuals to the best-fit model, followed by a pointwise multiplication of the best-fit baseline function;
\[
\vec y_{p} = (\vec m_{bf} + \vec r_{p}) \cdot (\mathbf{X} \hat\beta)
\]
The decorrelation and light curve fitting is done as before, to determine $\rho$. The procedure is repeated 1000 times with random perturbations to the shift in residuals, each time varying the starting value for $\rho$ to ensure the starting parameters do not affect the results. The resulting distribution of $\rho$ is then used to estimate its uncertainty in each wavelength channel.

\subsection[]{Results}
\label{sect:HD189733_results}

\begin{figure}
\includegraphics[width=84mm]{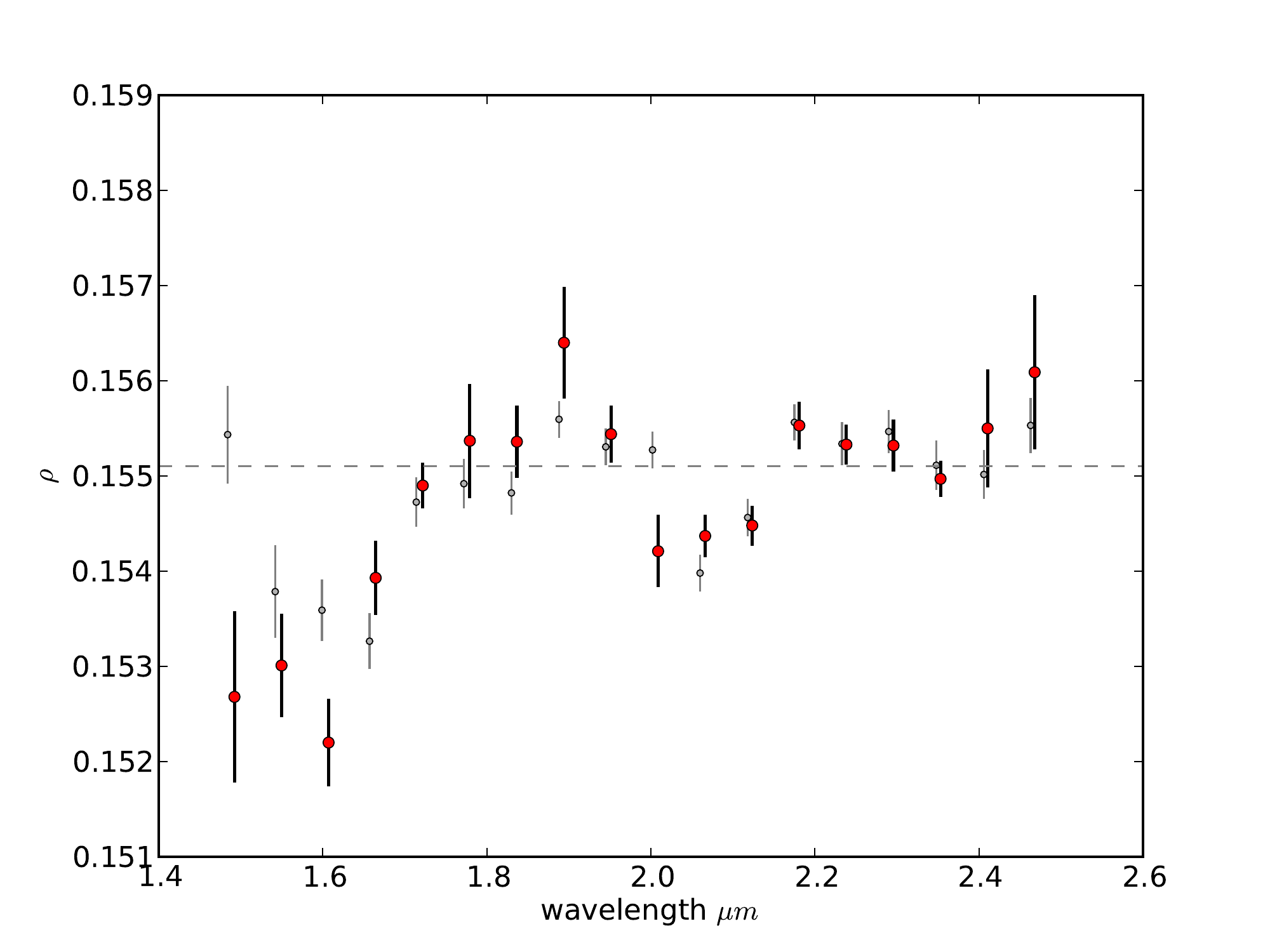}
\caption{Transmission spectrum of HD 189733 generated by determining the planet to star radius ratio from the decorrelated light curves for the 18 wavelengths channels. Our results are the red points, and those from S08 are shown in grey for comparison. The horizontal dashed line is the radius ratio measured for the white light curve.}
\label{fig:HD189733_trans_spec_all}
\end{figure}

The resulting NICMOS transmission spectrum for HD 189733 is shown in Fig.~\ref{fig:HD189733_trans_spec_all}. The data from S08 are also plotted for comparison, after converting from transit depth to $\rho$. For the most part, the spectra show the same basic shape. It is not clear exactly where the discrepancies in a few of the wavelength channels arise, but they are likely explained by one or a combination of the following; different pixel columns and widths used for the wavelength channels, different methods used to determine the background, the fact that we fit for the light curve rather than just taking an average of the in-transit orbit, that the decorrelation parameters are extracted slightly differently, and finally the corrections applied by S08 for limb darkening and star spots. We were unable to reproduce exactly the same results as S08 using a global background correction. Another obvious difference between the two spectra is that the uncertainties we calculate using the residual permutation method are significantly larger than those given in S08, particularly at the edge of the spectrum.

\begin{figure}
\includegraphics[width=84mm]{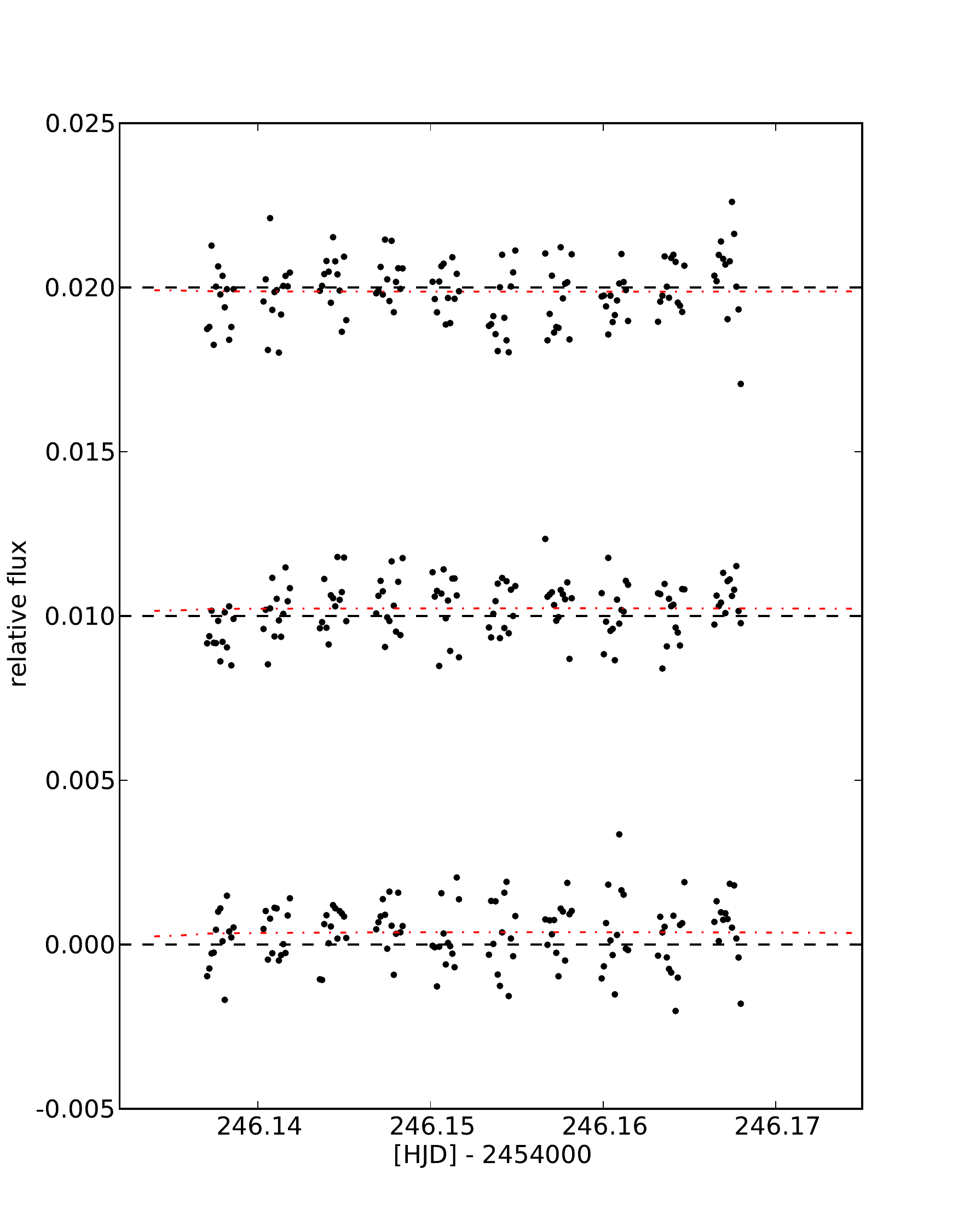}
\caption{Residuals of the in-transit orbit from a `constant radius' model generated from the white light curve planet-to-star radius ratio for three wavelength channels; 2.41, 2.07 and 1.66 \micron~from top to bottom, respectively. The limb darkening coefficients used were specific to each wavelength channel. The horizontal black dashed line shows the constant radius model, and the dashed red line the difference between it and the best-fit model determined when fitting $\rho$ for each wavelength channel. It is clear that systematic noise is present in the in-transit orbit for the three channels, at a level comparable to or even larger than the deviations from the constant radius. A similar level of systematics can be seen in all wavelength channels.}
\label{fig:HD189733_zoomed_residuals}
\end{figure}

However, we do not believe the residual permutation method fully accounts for the uncertainties in $\rho$. Fig.~\ref{fig:HD189733_zoomed_residuals} shows a plot of the in-transit residuals for three wavelength channels. The residuals are of decorrelated light curves, after subtracting a model generated from the planet-to-star radius ratio measured from the white light curve, and using limb darkening parameters specific to each wavelength channel. The difference between models generated for the best-fit planet-to-star radius ratio for each channel (i.e. the value in the transmission spectrum), and the planet-to-star radius ratio from the white light curve (i.e. a constant radius model) are also shown. As is clear for all three channels, systematic noise is present at a level comparable to or larger than the difference between the two models. The deviations from a constant transmission spectrum may therefore arise from systematics not removed from the in-transit orbit. A similar level of systematics is visible in the residuals for all wavelength channels.

\begin{figure}
\includegraphics[width=84mm]{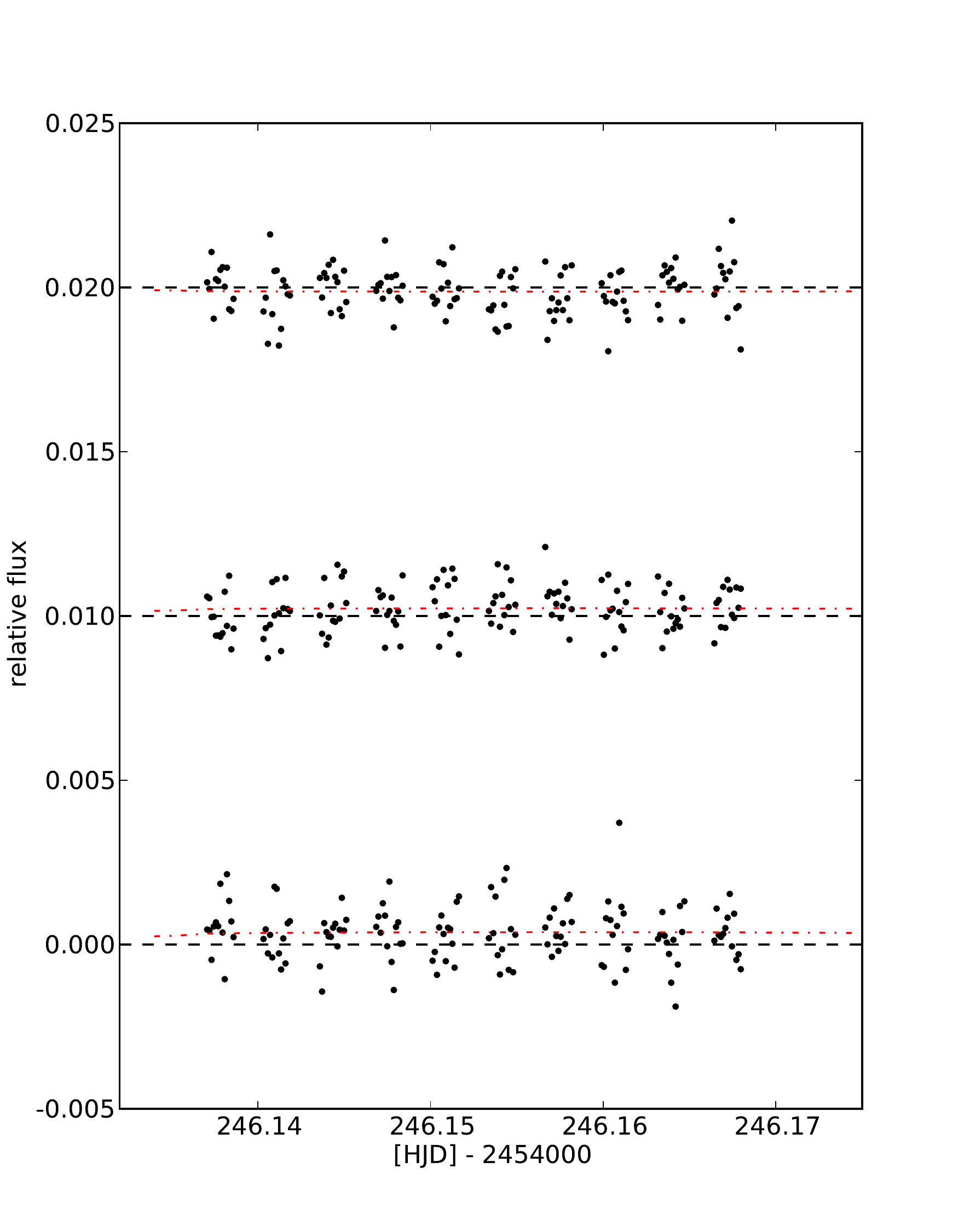}
\caption{Same as Fig.~\ref{fig:HD189733_zoomed_residuals} but after the channel-to-channel correction has been applied. In some cases the RMS of the residuals is reduced, but there are still systematics present that are comparable to or even larger than the deviations from a constant radius model.}
\label{fig:HD189733_zoomed_residuals_CTC}
\end{figure}

\begin{figure}
\includegraphics[width=84mm]{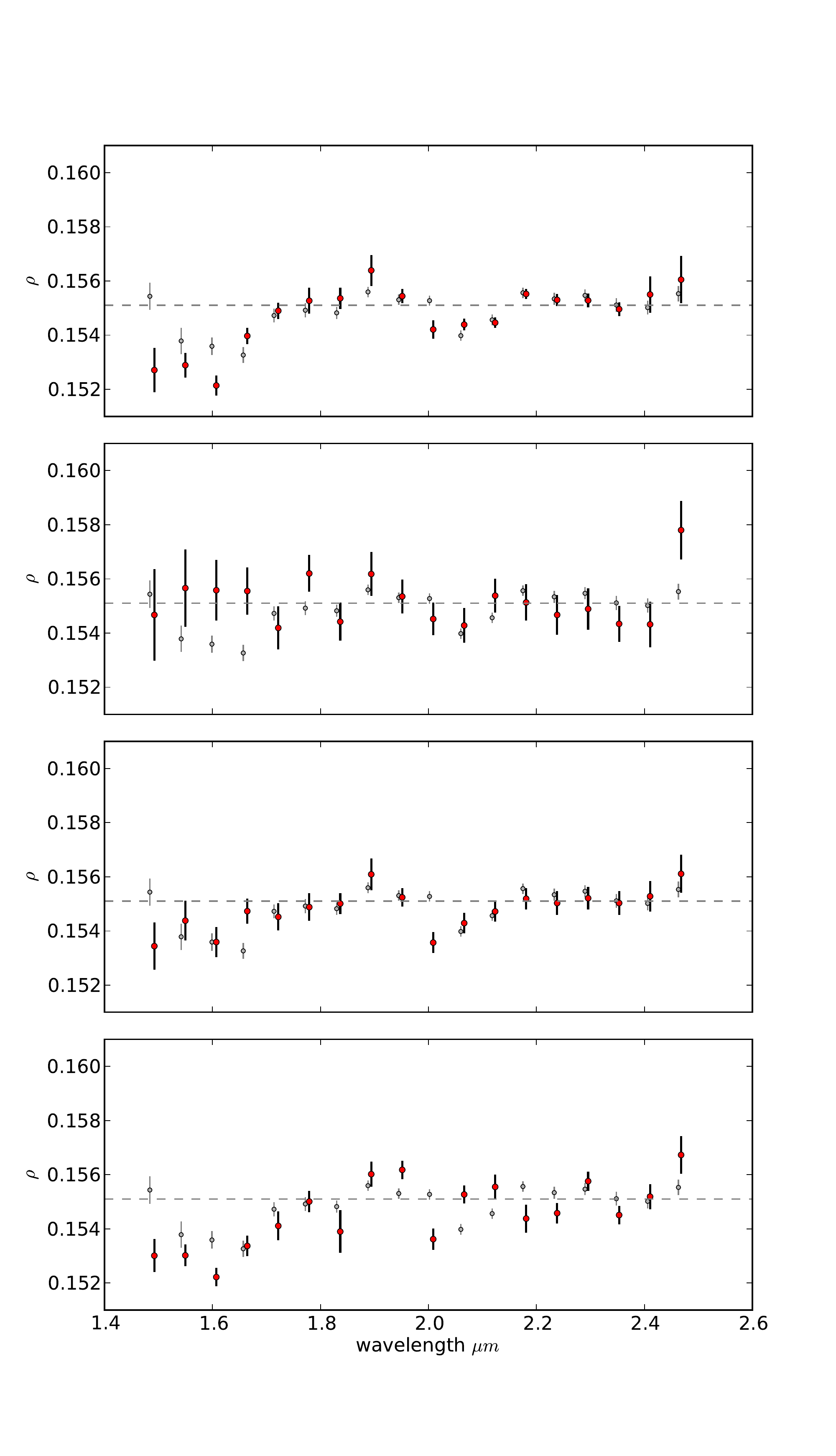}
\caption{Transmission spectra of HD 189733 generated by determining the planet radius $\rho$ for the 18 wavelengths channels. The top plot shows the transmission spectrum obtained after applying the channel-to-channel correction. The second plot shows the spectra by decorrelating the light curves only using orbits 2 and 4. The third plot shows the result of a `quadratic' decorrelation, and the bottom plot shows a linear decorrelation but with the angle vector ($\theta$) removed from the state matrix ($\mathbf{X}$). Again, our results are the red points, and those from S08 are shown in grey for comparison. The horizontal dashed grey line marks the level of the planet-to-star radius ratio of the white light curve. For the last three cases the spectra are altered considerably from the first, and none of the features reported in S08 are present in all spectra using the different methods.}
\label{fig:HD189733_trans_spec_var}
\end{figure}

We are yet to address one final correction on each wavelength channel carried out by S08, the `channel-to-channel' corrections. This involves taking the weighted average of the light curve residuals for all the wavelength channels, and subtracting them from individual channels after decorrelating the light curves. This should remove any common time-correlated systematics from the light curves. To check that this did not affect our results significantly, we carried out this step. Fig.~\ref{fig:HD189733_zoomed_residuals_CTC} shows the same light curves as Fig.~\ref{fig:HD189733_zoomed_residuals} after the channel-to-channel correction has been applied. This process does marginally reduce the RMS of the residuals, but does not completely remove the systematics. This is not surprising given all wavelength channels do not show the same structure of systematics. We would also not expect this to change the transmission spectrum, as applying the same correction to all light curves will only shift the depths uniformly. However, reducing the RMS may have an impact on the uncertainties. The resulting transmission spectrum is shown at the top of Fig.~\ref{fig:HD189733_trans_spec_var}. This does not cause significant changes in the transmission spectrum, or even in the calculated uncertainties. This indicates that the uncertainties in the transmission spectrum are probably dominated by determination of the baseline function. Recently, \citet{Burke_2010} identified systematic effects that should be accounted for with NICMOS grism spectroscopy, in particular `gain-like' variations that arise from seven states of the detector electronics. Importantly, they noted that the channel-to-channel correction would likely account for this.

\begin{figure}
\includegraphics[width=84mm]{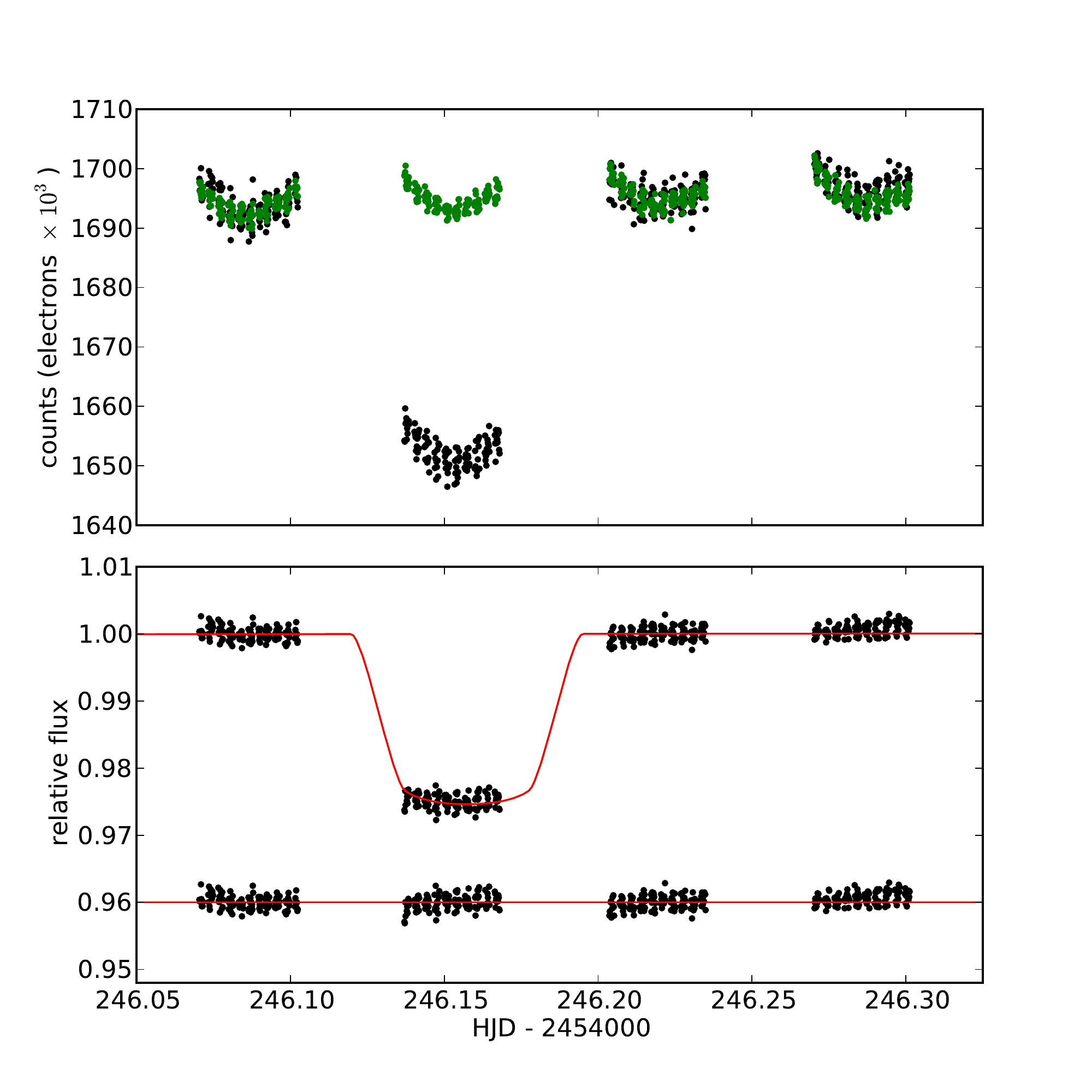}
\caption{Example of the decorrelation procedure on one of the HD 189733 wavelength channels, similar to Fig.~\ref{fig:HD189733_decorr_eg}, but showing a different wavelength channel, and now decorrelating only using orbits 2 and 4. It is clear that the decorrelation does not properly correct for orbit 5, and an artificial offset is introduced by the linear baseline model. Note that this offset is larger than the features in the reported transmission spectrum. We argue that the decorrelation process may introduce similar spurious offsets to the in-transit orbit, indistinguishable from a real atmospheric signature.}
\label{fig:HD189733_decorr_eg_24}
\end{figure}

Significant uncertainties in the depth will also come from offsets in the baseline flux level between the orbits. Offsets in the flux levels are seen in the raw light curves, and could easily be induced by the decorrelation model (as well as corrected for), as the decorrelation parameters themselves are seen to have offsets between orbits, therefore clearly a linear baseline model can produce `artificial' offsets. These flux offsets were also identified by \citet{Carter_2009}, who tried to fit decorrelation models based on state parameters to attempt to remove the flux offsets, and concluded the procedure was unreliable with no physical justification for a linear baseline model. As variations in the flux level for the in-transit orbit are fitted for when determining the planetary radius, the residual permutation algorithm cannot fully take this additional uncertainty into account, as the residuals will not contain the signal of any induced offset. To test for the possibility that the decorrelation methods may not fully correct for the flux level, we repeated the decorrelation process only using orbits 2 and 4 to determine the decorrelation coefficients. Referring to Fig.~\ref{fig:HD189733_decorr_parameters}, we are still interpolating for the in-transit orbit for all the decorrelation parameters. Therefore, if the model for the baseline flux is correct, we should still see similar results. Fig.~\ref{fig:HD189733_decorr_eg_24} shows an example of this decorrelation process. It is clear that orbit 5 has a significant offset from the baseline flux of orbits 2 and 4. This shows that the linear decorrelation method is not robust when extrapolating over the decorrelation parameters. If the functions describing the baseline function are \emph{not} linear in nature (which is by no means clear, as even assuming the baseline flux can realistically be written in terms of the optical state parameters is a strong assumption), a spurious offset could just as easily be induced in the in-transit orbit, leading to channel dependent transit depths, and resulting in a variable transmission spectrum.

The transmission spectrum obtained when decorrelating only using orbits 2 and 4 is shown in Fig.~\ref{fig:HD189733_trans_spec_var}. As expected, the uncertainties are now significantly larger than before (orbit 5 is not used in the fitting process). However, the transmission spectrum shows a significantly different structure than before (c.f. Fig~\ref{fig:HD189733_trans_spec_all}). In fact the only `feature' common to both, is the dip at around 2.1\,\micron, although its depth relative to the uncertainties varies. The edges of the spectrum show quite different behaviour. We interpret the significant differences in the two spectra as evidence that the linear baseline model is not sufficient to correct the light curves, and can induce unwanted offsets in the in-transit orbit, which are indistinguishable from a real atmospheric signal.

We also conducted additional tests to see how dependent the transmission spectrum is on the decorrelation model. A `quadratic' decorrelation was tried by adding the terms $\Delta X^2$, $\Delta Y^2$, $W^2$, $\theta^2$ and $T^2$, to the state matrix ($\phi_H$ already has a quadratic term). We also made tests decorrelating the spectra after removing each vector from the state matrix. We found that the most important decorrelation parameter to be $\theta$, and to a lesser extent $\Delta Y$. The other decorrelation parameters play a much less important role. This is consistent with the findings in S08, and is rather obvious after closer inspection of the decorrelation parameters in Fig.~\ref{fig:HD189733_decorr_parameters}. $\theta$ and $\Delta Y$ show significant offsets between the in-transit orbit and the out-of-transit orbits. Thus, decorrelating with these state parameters causes larger shifts in the in-transit orbit than the others, and consequently the most significant corrections to the transmission spectrum.

The spectra produced using the `quadratic' decorrelation, and after removing $\theta$ from the state matrix are shown in Fig.~\ref{fig:HD189733_trans_spec_var}. Note all of the spectra plotted in Fig.~\ref{fig:HD189733_trans_spec_var} are after the channel-to-channel correction has been applied, but its exclusion does not alter the results significantly for any case. It is clear from these plots that the various techniques used to decorrelate the light curves produce quite different output transmission spectra. Each of the features reported in S08 vanishes in at least one of the cases. As $\theta$ was previously shown to be the most important parameter in the decorrelation, the interpretation of molecular features in the spectrum is heavily dependant on a linear dependance of the flux level with the angle the spectrum makes on the detector. We do not argue that any of these methods are better than the others or the method used by S08, but they give incompatible results and there is no way to distinguish between them. Whilst most of the systematic effects in the light curves are caused by intra-pixel sensitivity changes as the spectral trace drifts and rotates on the detector, it is not clear that a linear decorrelation will remove these systematics. In fact, we have shown that the decorrelation can cause spurious offsets in the in-transit flux level.

Furthermore, the largest disagreements between the spectra occur at the edges, where the background is expected to significantly affect the transit depth. The ratio of the true transit depth ($D$) is related to the measured transit depth ($D^\prime$) by
\[
\frac{D}{D^\prime} = 1 + \frac{B}{F_0},
\]
where $B$ is the (uncorrected) background counts, and $F_0$ the out-of-transit counts. Therefore, an underestimated or overestimated background will lead to incorrect transit depths. This effect scales with the ratio of the background to the out-of-transit flux. Thus where the flux count is lower, i.e. at the fainter channels (the edges) of the spectrum, this effect will be much greater. For the faintest channels, we have approximately $150\,000 \times 5$ electrons (Fig.~\ref{fig:HD189733_1D_spectra}) collected over 35 $\times$ 5 pixels. If we assume a background correction error of 50 electrons per pixel (a conservative estimate considering the background can vary by up to 200 electrons per pixel), the ratio of true depth to measured depth is about 1.012, more than enough to significantly affect the measured transit depth. Clearly we cannot trust the edges of the spectra to give us an accurate measurement of the planet-to-star radius ratio.

\section{GJ-436}
\label{sect:GJ436}

\subsection{Observations}
\label{sect:GJ436_obs}

A transit of GJ-436 was monitored on 10-11th November 2007 with HST/NICMOS, using the G141 grism covering the wavelength range 1.1 -- 1.9 \micron. Analyses of these data were first reported in \citet[][hereafter P09]{Pont_2009}. We only used the first of two light curves reported in P09 for this work, as they concluded that the noise and systematic effects were lower during the first visit. GJ-436 is again not in the continuous viewing zone of the HST, and the transit was observed over four half-orbits, consisting of 935 spectra in total, all with exposure times of 1.993 seconds. The first, second and fourth orbits cover the out-of-transit part of the light curve and consist of 180, 252 and 251 observations, respectively, and are used to determine the photometric baseline. The third orbit was taken in-transit, covering mid-transit and egress, and consists of the remaining 252 images. Similarly to HD 189733, some exposures were taken at the beginning of the first orbit to set the wavelength calibration.

Again, calibrated images include all basic calibrations except for flat-fielding. The background with the G141 grism is much lower than with the G206 grism, and is therefore the background correction is less of an issue for this data set. The background was again very stable over the duration of the observations, with a typical level of only $25-30$ electrons per pixel. The spatial variation over the detector is about 10 electrons per pixel. We again tried flat-fielding using a flat-field taken with the G141 grism in place, but this did not provide a satisfactory correction and we did not use the flat-field corrected images for this analysis.

\begin{figure}
\includegraphics[width=84mm]{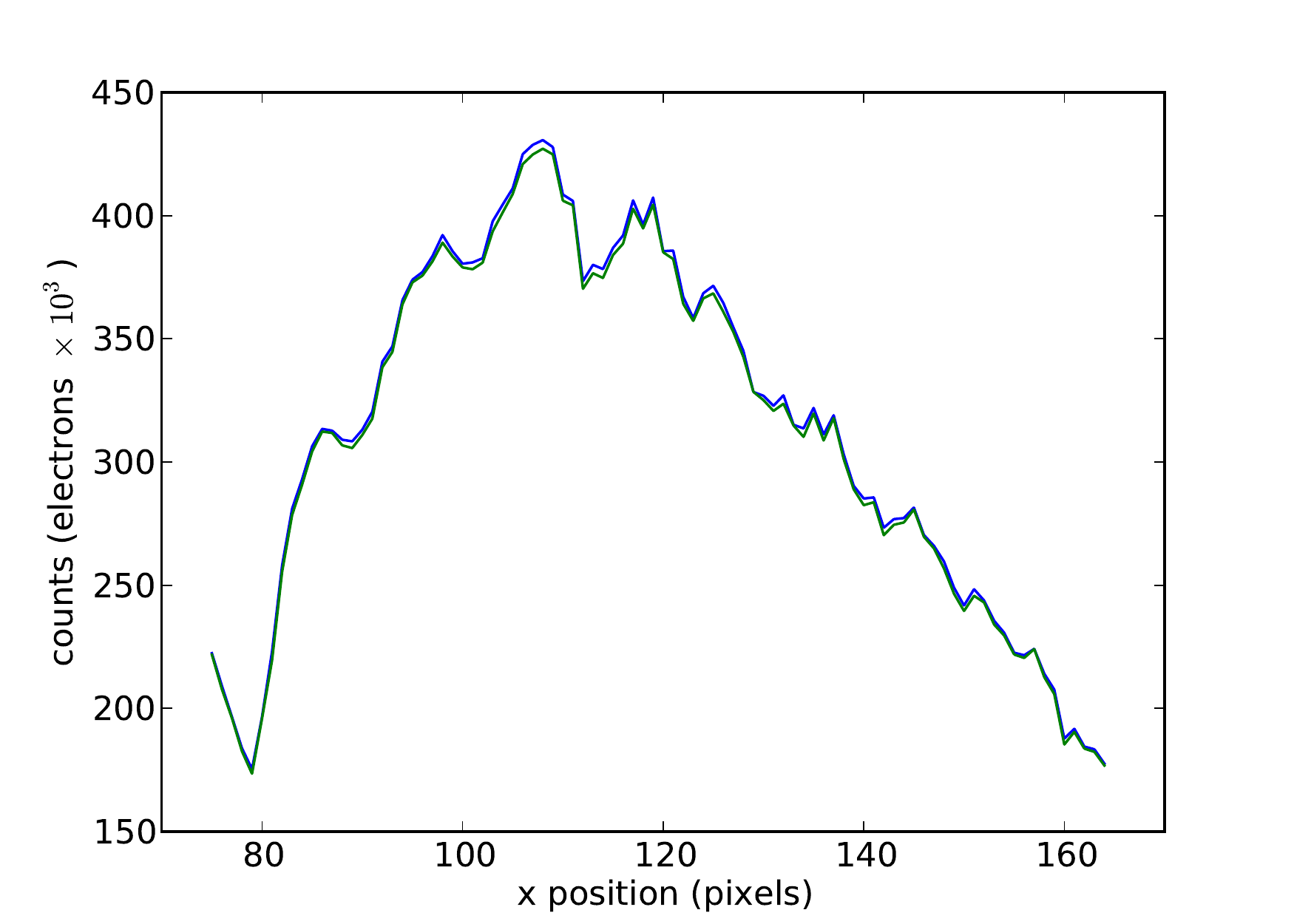}
\caption{Extracted 1D spectra for GJ-436 of a typical in-transit (green) and out-of-transit (blue) observation showing the number of electrons collected per pixel channel. The wavelength decreases along the x-axis.}
\label{fig:GJ436_1D_spectra}
\end{figure}

\begin{figure}
\includegraphics[width=84mm]{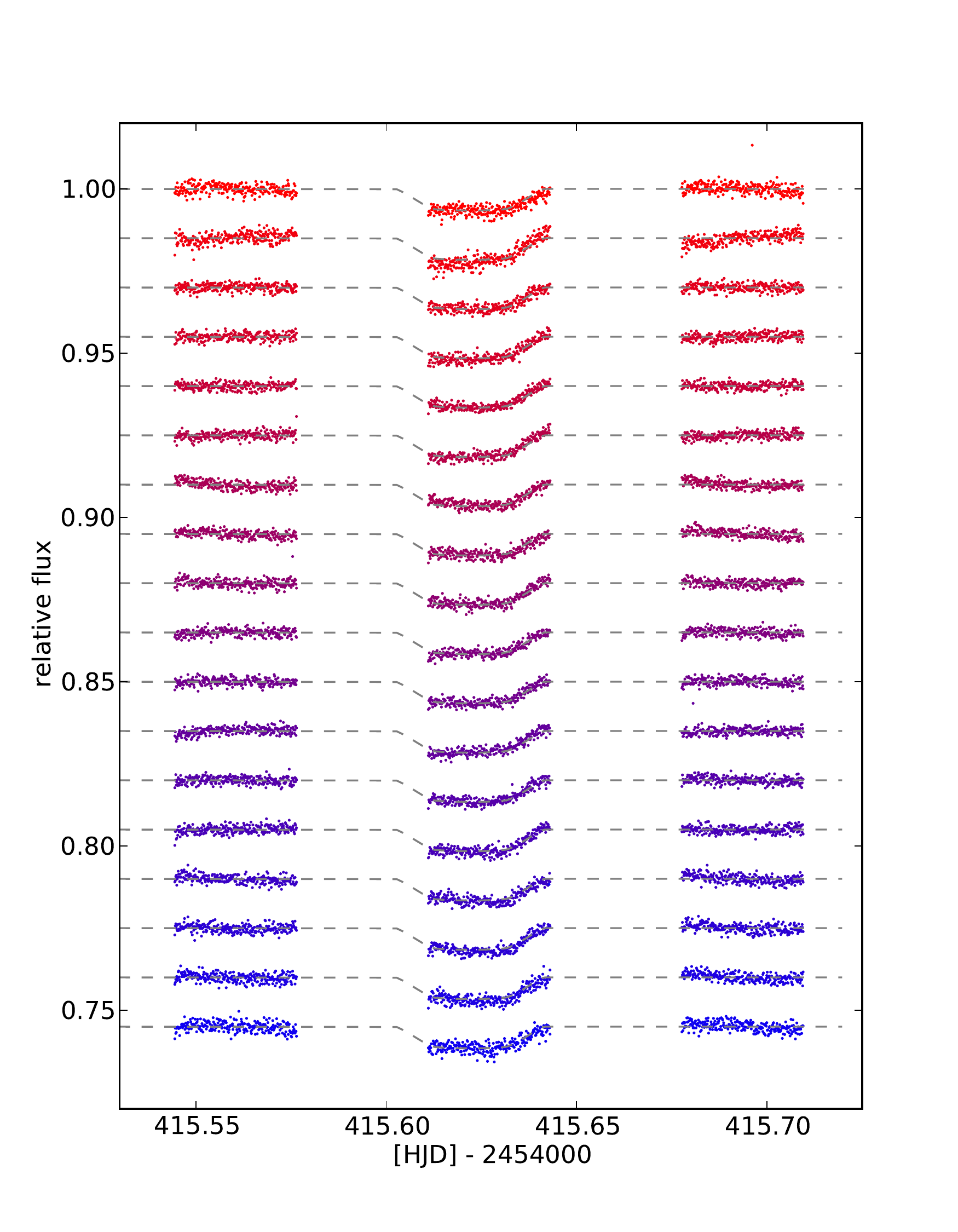}
\caption{Raw light curves for each of the 18 wavelength channels for GJ-436,  from 1.87\,\micron~(top) to 1.19\,\micron~(bottom), after normalising each one by fitting a linear function through orbits 2 and 4. Similarly to HD 189733, correlated noise is seen in each light curve.}
\label{fig:GJ436_normalised_lightcurves}
\end{figure}

\begin{figure}
\includegraphics[width=84mm]{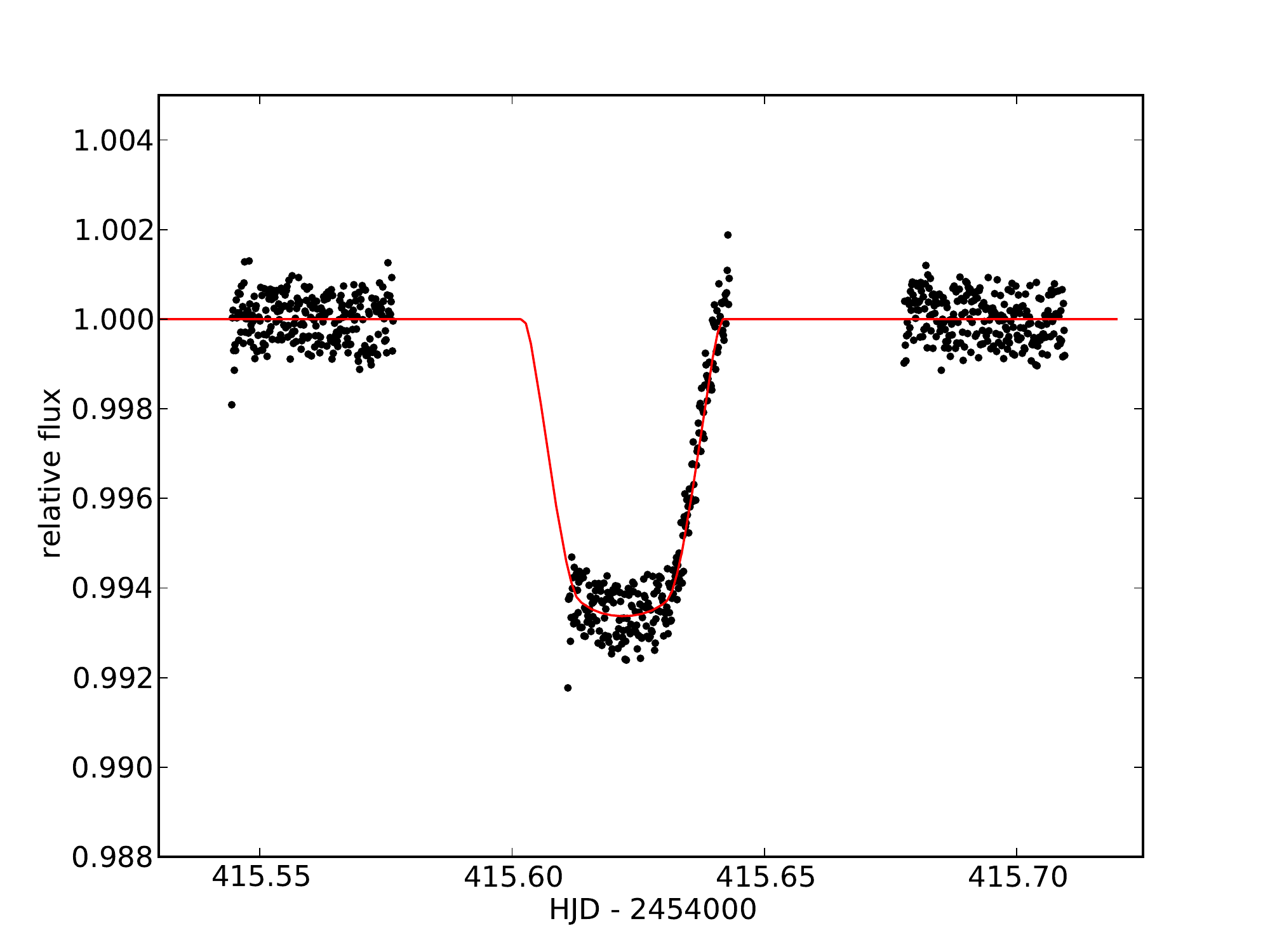}
\caption{Raw `white' light curve of GJ-436 found by integrating the flux from each spectra over all wavelengths, showing the sampling of the GJ-436 transit, excluding the first orbit.}
\label{fig:GJ436_wlc}
\end{figure}

\begin{figure}
\includegraphics[width=84mm]{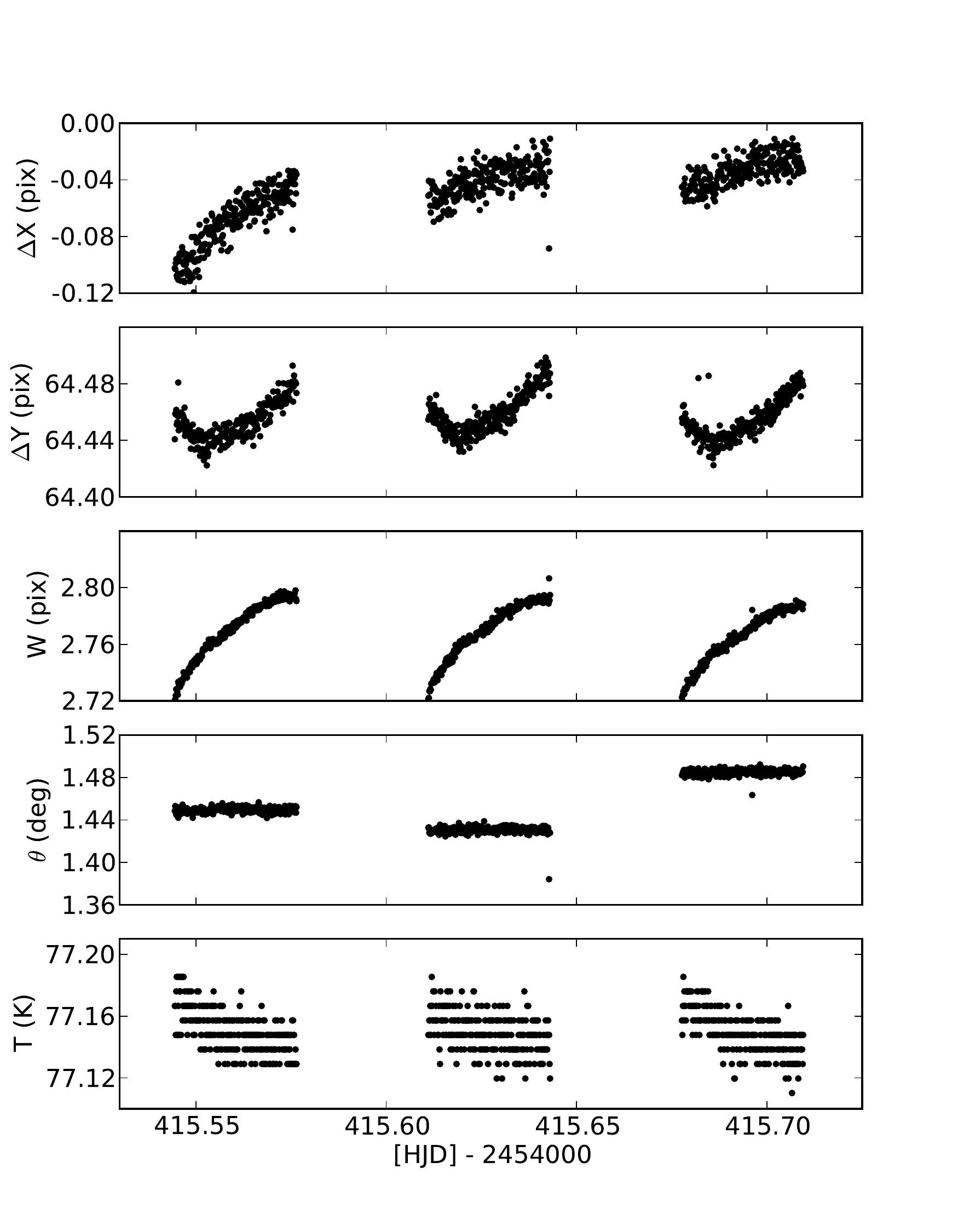}
\caption{Extracted de-correlation parameters for GJ-436 during orbits 2--4 as a function of time. For the angle of the spectral trace $\theta$, the baseline function must be extrapolated for the in-transit orbit, hence this parameter cannot be used in the decorrelation process, without introducing large offsets in the transit light curve.}
\label{fig:GJ436_decorr_parameters}
\end{figure}

The spectra and light curves were extracted using the same technique as for HD 189733. The spectra were extracted along 90 pixel columns, with a width of 20 pixels. The width of 20 pixels was selected to minimise the RMS in orbits 2 and 4 of the white light curve. Again we experimented using both global and column-specific background corrections, but it made insignificant difference due to the low background. Fig.\ref{fig:GJ436_1D_spectra} shows a 1D spectrum for an in-transit and out-of-transit observation. About 420\,000 electrons were collected per exposure in the brightest pixel channel, and approximately 170\,000 electrons in the faintest pixel channel. The raw light curves were then constructed as for HD 189733 by binning the spectra into 5 pixel bins, and are shown in Fig.~\ref{fig:GJ436_normalised_lightcurves}. They show similar systematic effects to HD 189733. The white light curve was extracted using 110 pixel columns and is shown in Fig.~\ref{fig:GJ436_wlc}, excluding orbit 1 (left out of the remaining analysis due to large systematics), and the optical state parameters are shown in Fig.~\ref{fig:GJ436_decorr_parameters}.

\subsection{Analysis}
\label{sect:GJ436_analysis}

\begin{figure}
\includegraphics[width=84mm]{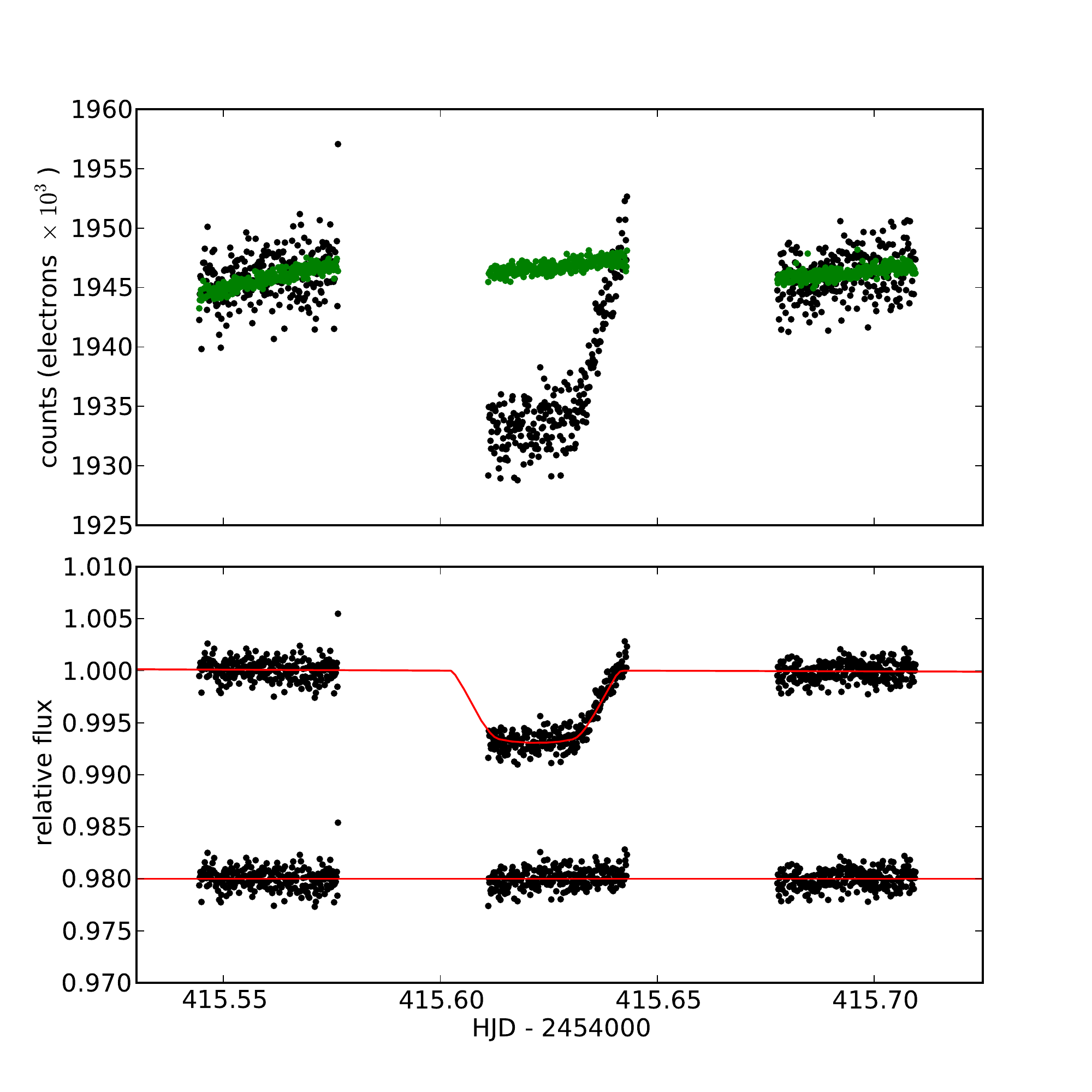}
\caption{Example of the decorrelation procedure on one of the GJ-436 wavelength channels, showing the raw and decorrelated light curve at the top and bottom, respectively. The green points represent the baseline function.}
\label{fig:GJ436_lcv1_decorr_eg}
\end{figure}

Each light curve was decorrelated using the procedure described in Sect.~\ref{sect:HD189733_analysis}. The decorrelation coefficients were determined for orbits 2 and 4. Referring to Fig.~\ref{fig:GJ436_decorr_parameters}, it is clear that we are extrapolating for the angle of the spectrum when correcting the in-transit orbit. We therefore must exclude this parameter from the multi-linear decorrelation, otherwise this results in a spectrum with very large features caused by spurious offsets in the in-transit baseline function. An example of the decorrelation procedure is shown in Fig.~\ref{fig:GJ436_lcv1_decorr_eg}.

\begin{figure}
\includegraphics[width=84mm]{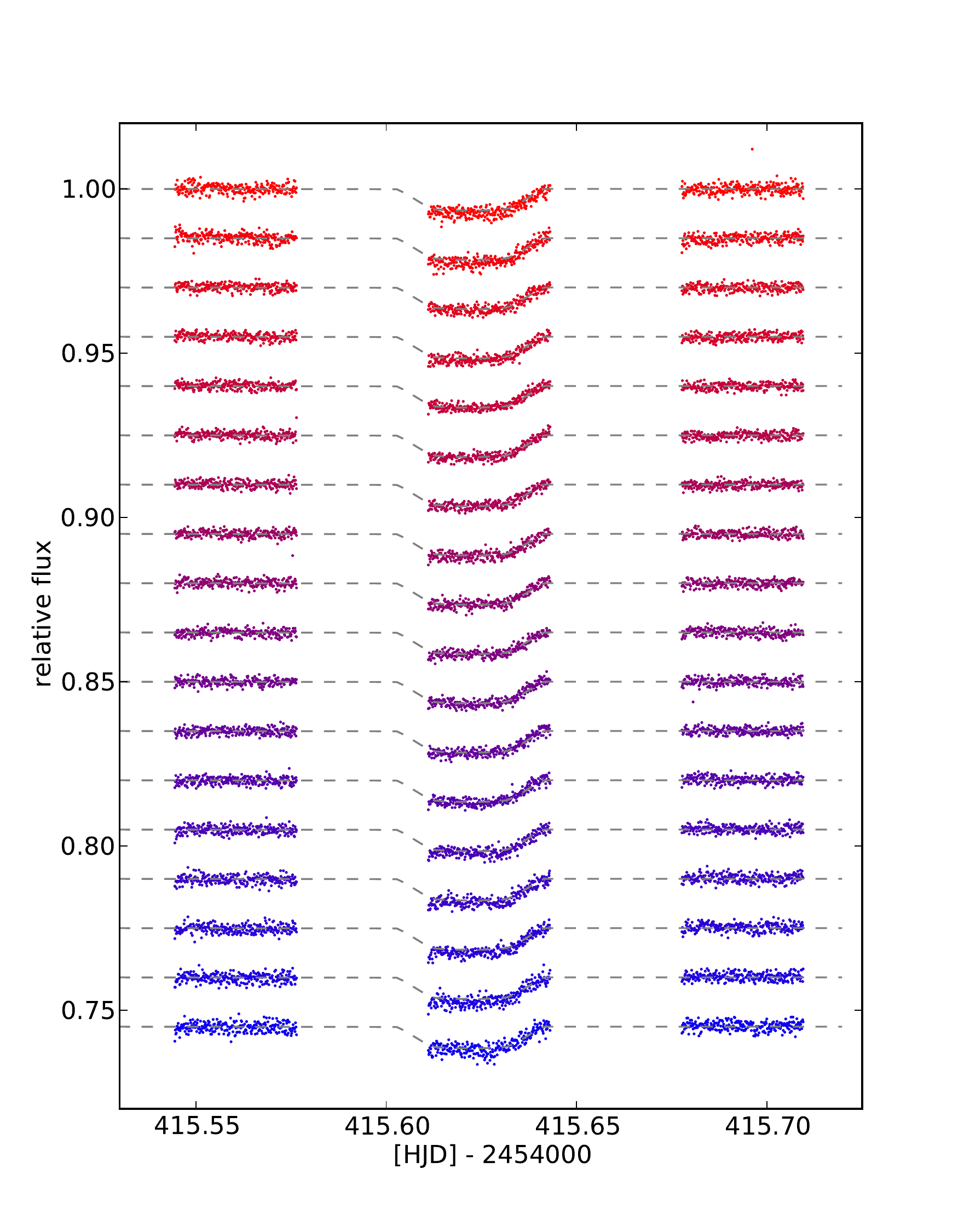}
\caption{De-correlated transit light curves of GJ-436 for each of the 18 wavelength channels. Again, much of the time-correlated systematic noise is removed from the light curves, but some remains.}
\label{fig:GJ436_decorr_lightcurves}
\end{figure}

The decorrelated light curves are shown in Fig.~\ref{fig:GJ436_decorr_lightcurves}. Each of the light curves was then fitted using the transit models described in Sect.~\ref{sect:HD189733_analysis} to determine $\rho$ as a function of wavelength, fixing the stellar and orbital parameters at those determined by P09, who undertook a thorough analysis of the integrated light curve. The central transit time was determined from the white light curve (as a well sampled egress allows an accurate determination of the central transit time), and fixed at this value when fitting for each wavelength channel. The limb darkening parameters were calculated using the method of \citet{Sing_2010} as for HD 189733, and the uncertainties were calculated using the residual permutation algorithm described previously, with 1000 light curves generated with random shifts applied to the light curve residuals.

\subsection{Results}

\begin{figure}
\includegraphics[width=84mm]{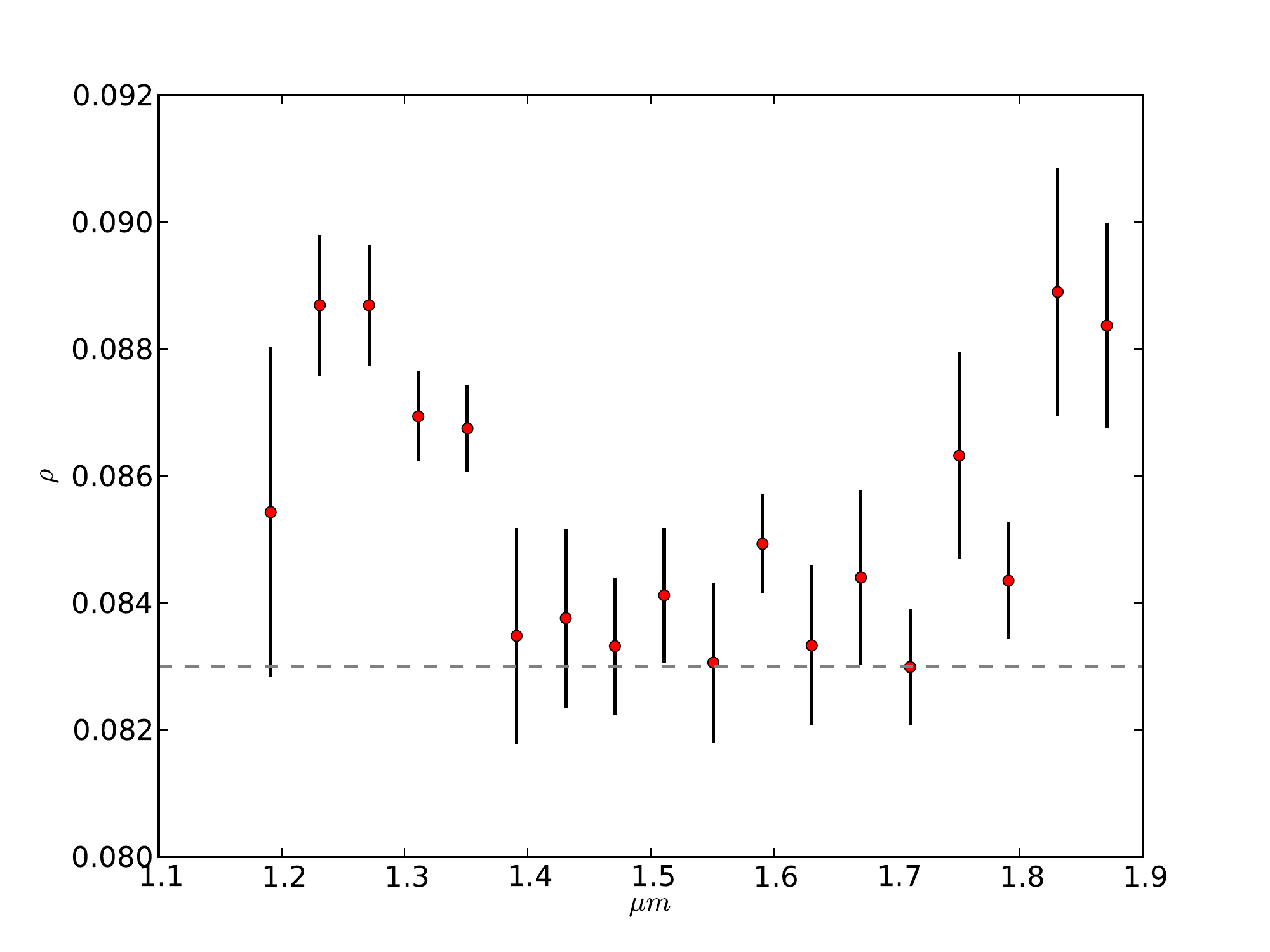}
\caption{Transmission spectrum of GJ-436 generated by determining the planet-to-star radius ratio from the decorrelated light curves for the 18 wavelengths channels. The horizontal dashed line shows the planet-to-star radius ratio calculated from the white light curve.}
\label{fig:GJ436_trans_spec_all}
\end{figure}

The resulting transmission spectrum for GJ-436 is shown in Fig.~\ref{fig:GJ436_trans_spec_all}. There seems to be a significant variation of the radius ratio with wavelength. Unfortunately, with only two out-of-transit orbits to work with, we do not have enough data to test the decorrelation model with different orbits. The planet-to-star radius ratio is calculated as $\rho = 0.0830\pm0.0005$ for the white light curve\footnote{This uncertainty is obtained when only fitting for the transit depth and a linear baseline function, and therefore is likely underestimated.}, consistent with P09. The transmission spectrum does not appear consistent with this value, as nearly all wavelength channels in the transmission spectrum give a larger depth. However, the transmission spectrum is consistent with the white light curve if we decorrelate only using the orbital phase parameters. A likely explanation for this is that the linear decorrelation process induces false offsets in the in-transit flux level, as was suspected for HD 189733.

It is interesting that the amplitude of the measured transit depth ($\simeq\rho^2$) shows a similar amplitude to that of HD 189733. Any variations in the transmission spectrum is likely due to systematic noise in the detector that we have not accounted for in our error analysis, rather than any real physical effect. Perhaps this represents the limit of NICMOS transmission spectroscopy observations due to correlated noise, without a more robust way to remove or control the systematics.

\section{XO-1}
\label{sect:XO1}

\subsection{Observations}
\label{sect:XO1_obs}

A transit of XO-1 was observed on 21 February 2008 with HST/NICMOS, using the G141 grism. These data were first reported in \citet[][hereafter T10]{Tinetti_2010}. XO-1 is a similar spectral type to HD 189733, but it is considerably fainter, with a J-band magnitude of 9.94 (c.f. 6.07 for HD 189733). The transit was observed over five half orbits, consisting of 279 spectra in total, each with an exposure time of 40 seconds. The first, second and fifth orbit cover the out-of-transit part of the light curve, and consist of 56, 56 and 55 spectra, respectively. The third orbit covers the ingress with 56 spectra, and the fourth covers mid-transit, consisting of 55 spectra. XO-1 has a longer transit duration than HD 189733 or GJ-436, hence two consecutive orbits cover in-transit data. Again, some exposures were taken prior to the first orbit to set the position for wavelength calibration.

\begin{figure}
\includegraphics[width=84mm]{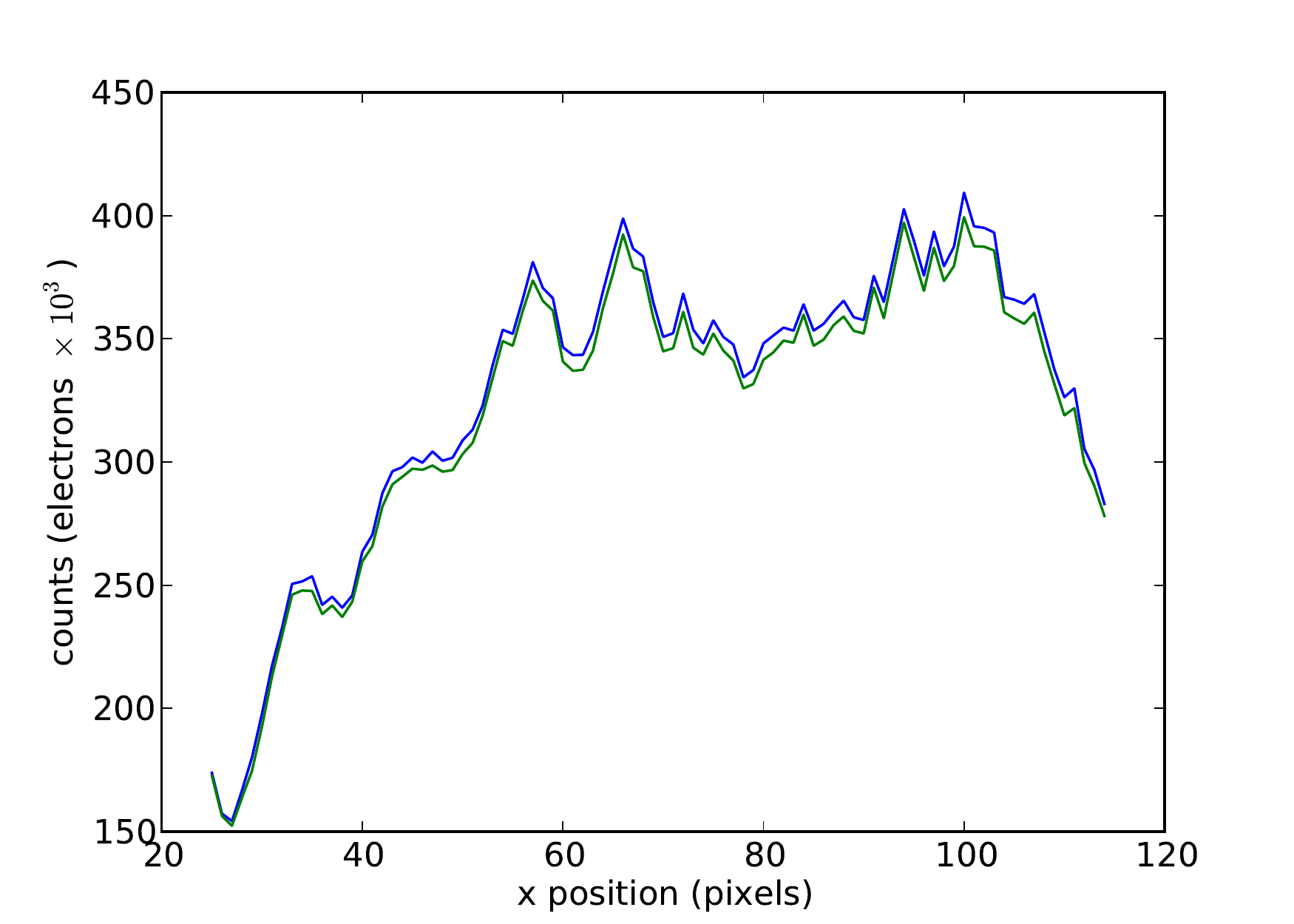}
\caption{Extracted 1D spectra for XO-1 of a typical in-transit (green) and out-of-transit (blue) observation showing the number of electrons collected per pixel channel.}
\label{fig:XO1_1D_spectra}
\end{figure}

\begin{figure}
\includegraphics[width=84mm]{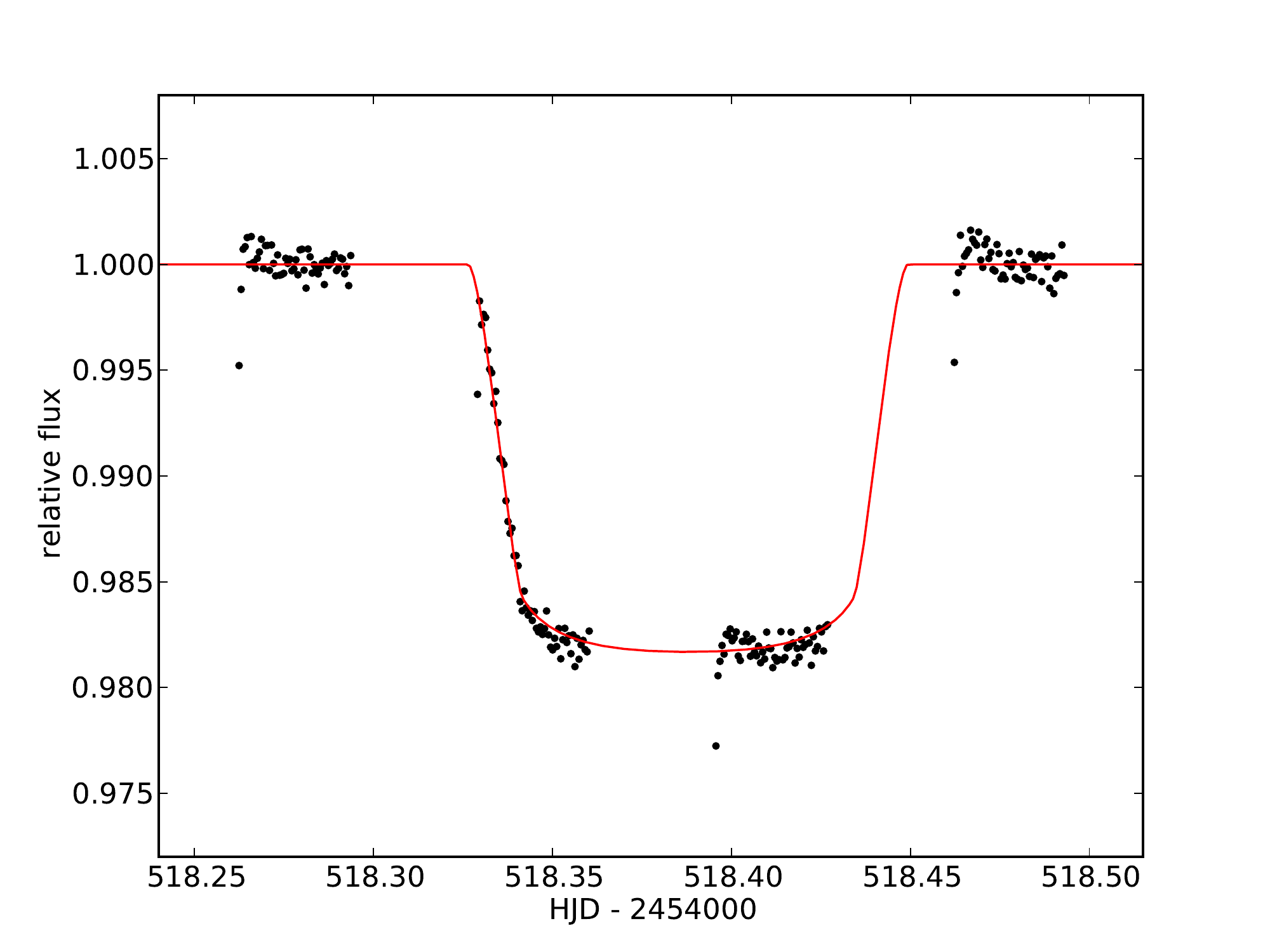}
\caption{Raw `white' light curve of XO-1 found by integrating the flux from each spectra over all wavelengths, showing the sampling of the XO-1 transit, excluding the first orbit.}
\label{fig:XO1_wlc}
\end{figure}

\begin{figure}
\includegraphics[width=84mm]{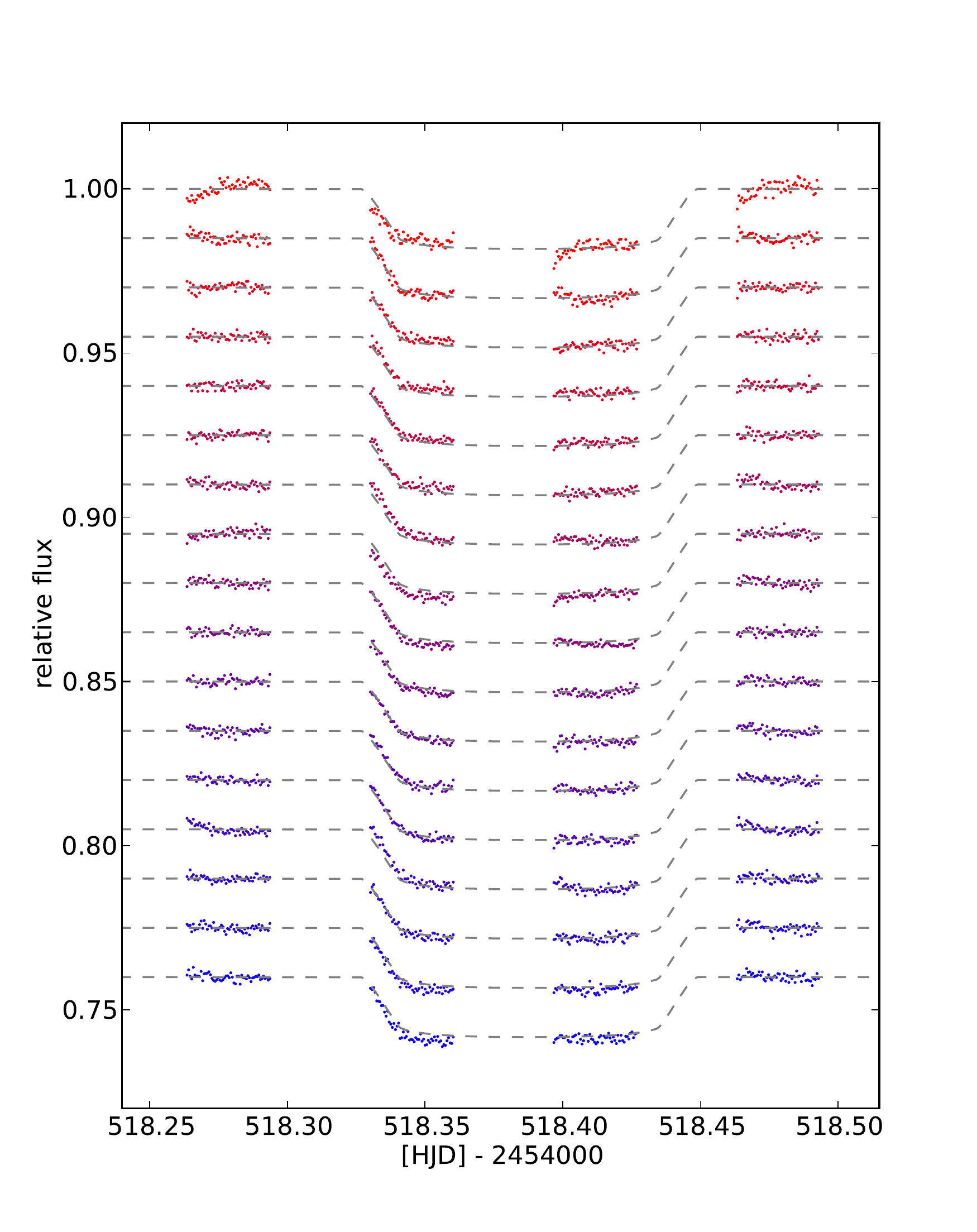}
\caption{Raw light curves of XO-1 for each of the 18 wavelength channels from 1.87\,\micron~(top) to 1.19\,\micron~(bottom) excluding orbit 1, and after normalising each one by fitting a linear function through orbits 2 and 5.}
\label{fig:XO1_normalised_lightcurves}
\end{figure}

The images were treated and the light curves extracted using the techniques described in Sects.~\ref{sect:HD189733_analysis} and \ref{sect:GJ436_analysis}. The 1D spectra were extracted along 90 pixel columns of width 16 pixels. The background was estimated from a column strip above each wavelength channel. We also experimented with flat-fielding and global background corrections, but neither significantly affects the final results. Fig.\ref{fig:XO1_1D_spectra} shows typical 1D spectra extracted for in-transit and out-of-transit observations, with the number of electrons collected per image per pixel column ranging from approximately \,160\,000 to \,400\,000. Fig.~\ref{fig:XO1_normalised_lightcurves} shows the light curves extracted for each of the 18 wavelength channels, after binning into 5 pixel bins. Fig.~\ref{fig:XO1_wlc} shows the resulting white light curve, extracted over 110 pixel columns. The first few points in each orbit exhibit strong variation in flux, and are thus excluded from the subsequent analysis.

\begin{figure}
\includegraphics[width=84mm]{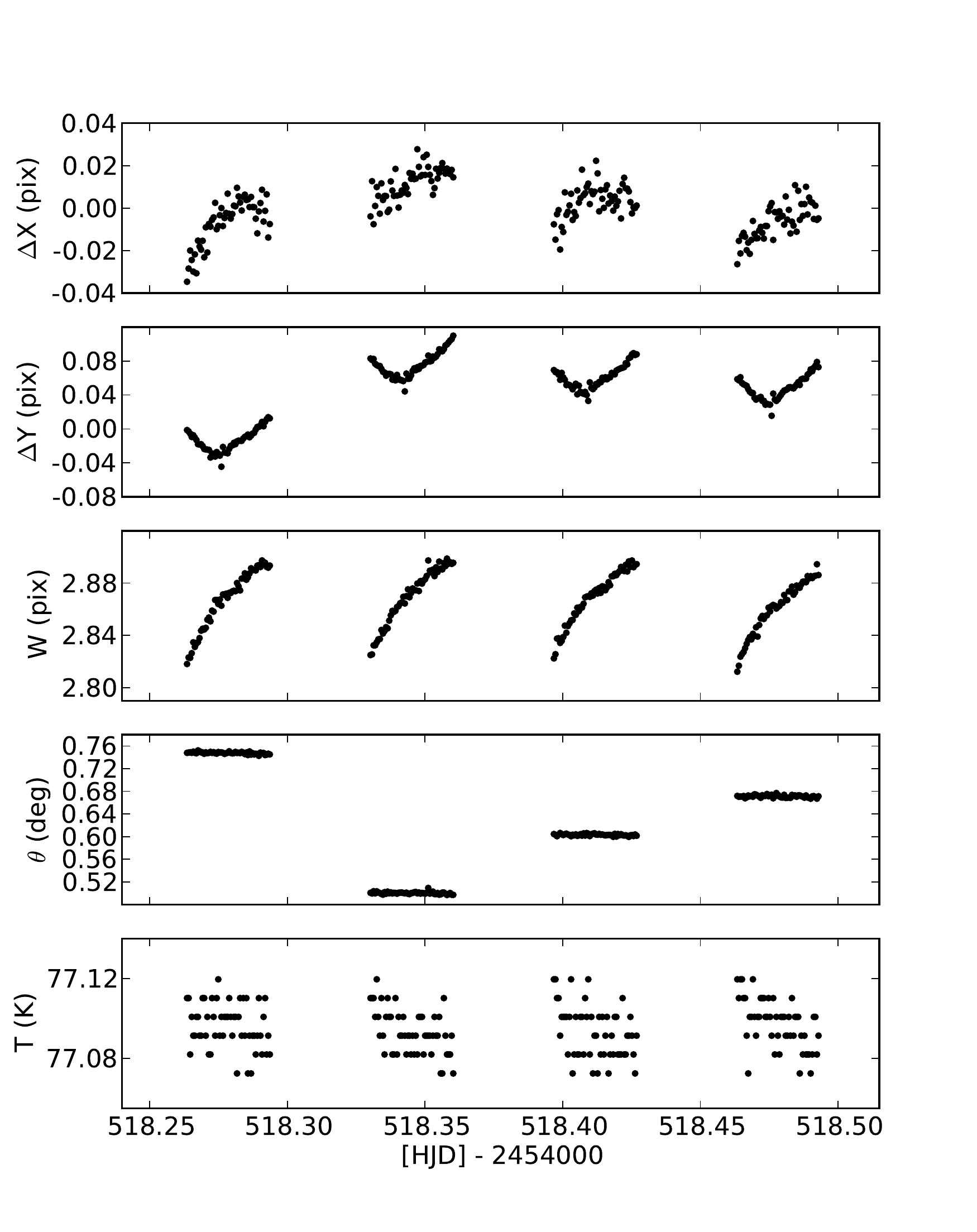}
\caption{Extracted optical state parameters for XO-1 during orbits 2--5. For the in-transit orbits (3 and 4), we have to extrapolate for $\Delta X$, $\Delta Y$ and $\theta$, making the decorrelation process extremely difficult.}
\label{fig:XO1_decorr_parameters}
\end{figure}

The white light curve shows much larger systematics than either the HD 189733 and GJ-436 light curves. Consequently, the light curves for each wavelength channel suffer from significant levels of correlated noise. The decorrelation parameters were extracted as for HD 189733 and GJ-436, and are plotted in Fig.~\ref{fig:XO1_decorr_parameters}.

\subsection{Analysis}

\begin{figure}
\includegraphics[width=84mm]{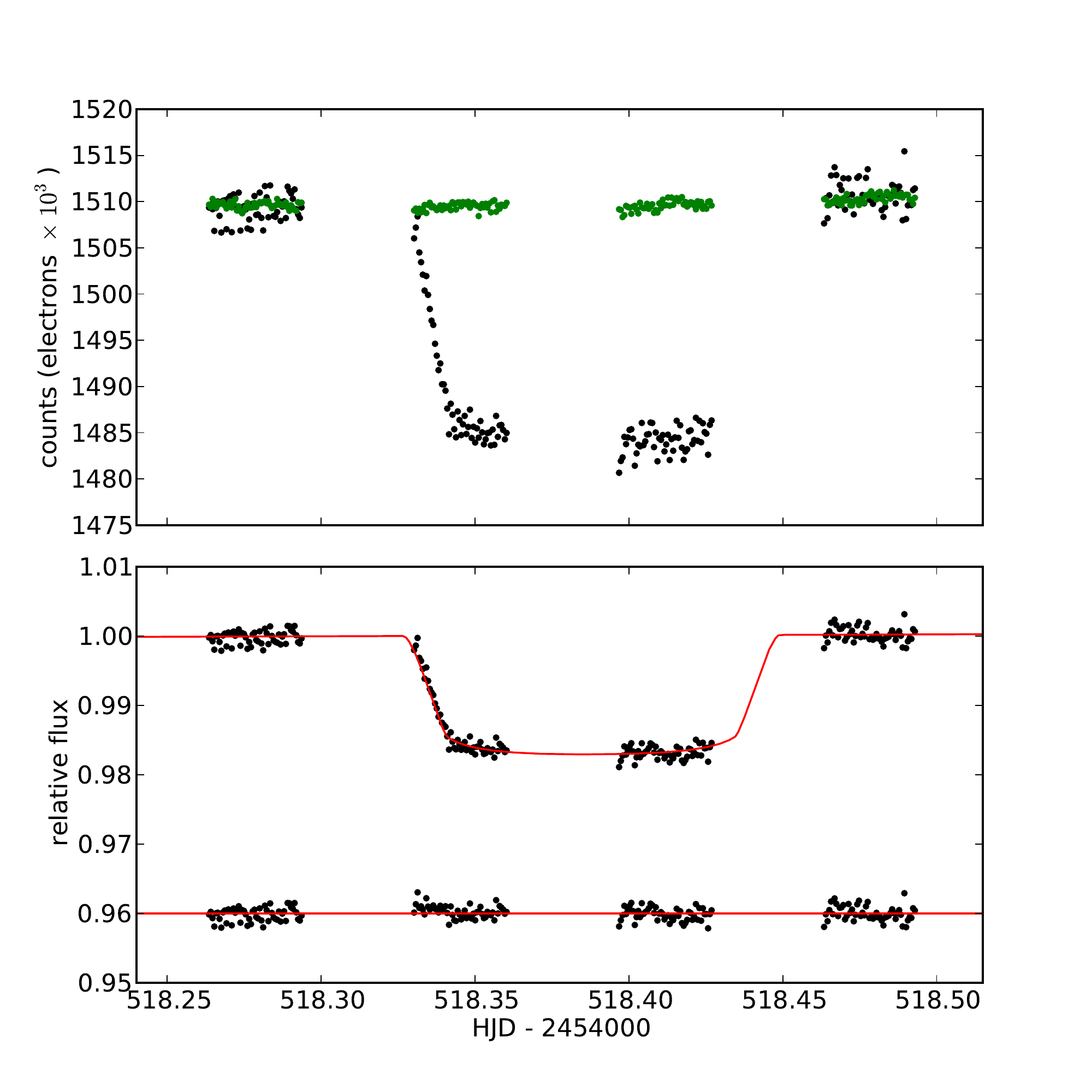}
\caption{Example of the decorrelation procedure on one of the XO-1 wavelength channels, showing the raw and decorrelated light curve at the top and bottom, repectively. The green points represent the baseline function. It is clear this does not provide a satisfactory correction for the systematics, which is obvious in the residuals for all orbits.}
\label{fig:XO1_decorr_eg}
\end{figure}

Given the large amount of systematics, one would like to apply the decorrelation used for HD 189733 and GJ-436 on these data. However, it is clear from Fig.~\ref{fig:XO1_decorr_parameters}, that for the in-transit orbits (3 and 4), we must extrapolate if we are to use corrections for $\Delta X$, $\Delta Y$ and $\theta$. We therefore must exclude these decorrelation parameters from the procedure. An example of the decorrelation procedure using only $W$, $T$, $\phi_H$ and $\phi_H^2$, is shown in Fig.~\ref{fig:XO1_decorr_eg}. The large residuals show the decorrelation process does not provide a more satisfactory correction.

\begin{figure}
\includegraphics[width=84mm]{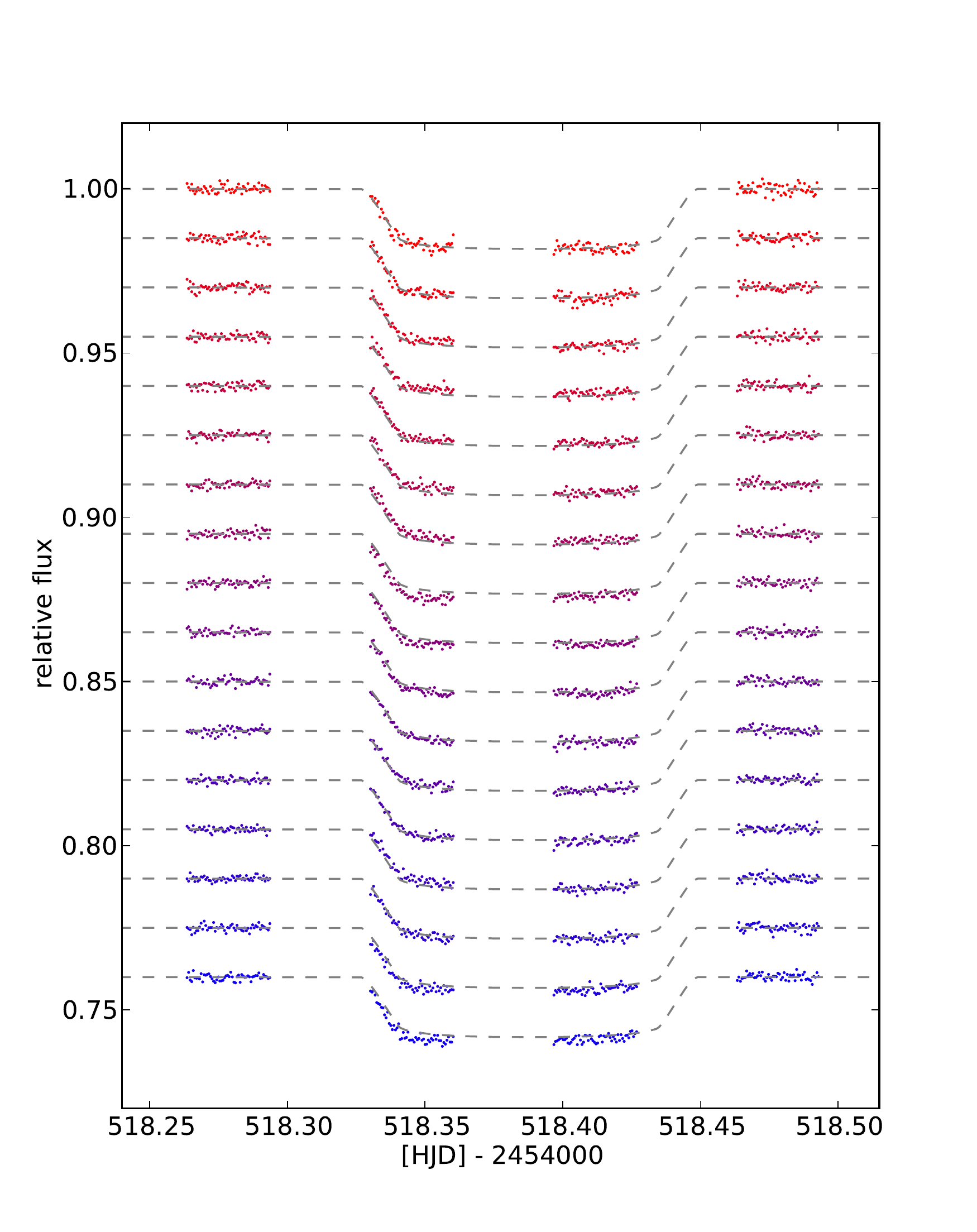}
\caption{De-correlated light curves of XO-1 for each of the 18 wavelength channels. Some of the time-correlated systematic noise is removed from the light curves; however, a significant amount remains which could not be removed in the decorrelation procedure with any linear combination of decorrelation parameters.}
\label{fig:XO1_decorr_lightcurves}
\end{figure}

The decorrelated light curves are shown in Fig.~\ref{fig:XO1_decorr_lightcurves}. Time-correlated noise still remains in the light curves, but no combination of decorrelation parameters produces a satisfactory result. The light curves were then fitted using the transit models described previously, fixing the orbital and stellar parameters at those determined by \citet{Torres_2008}. The central transit time was again set by fitting the white light curve with a transit model, and was then fixed when fitting the light curve for each wavelength channel. The uncertainties were determined using the residual permutation algorithm described, with 1000 realisations.

\subsection{Results}

\begin{figure}
\includegraphics[width=84mm]{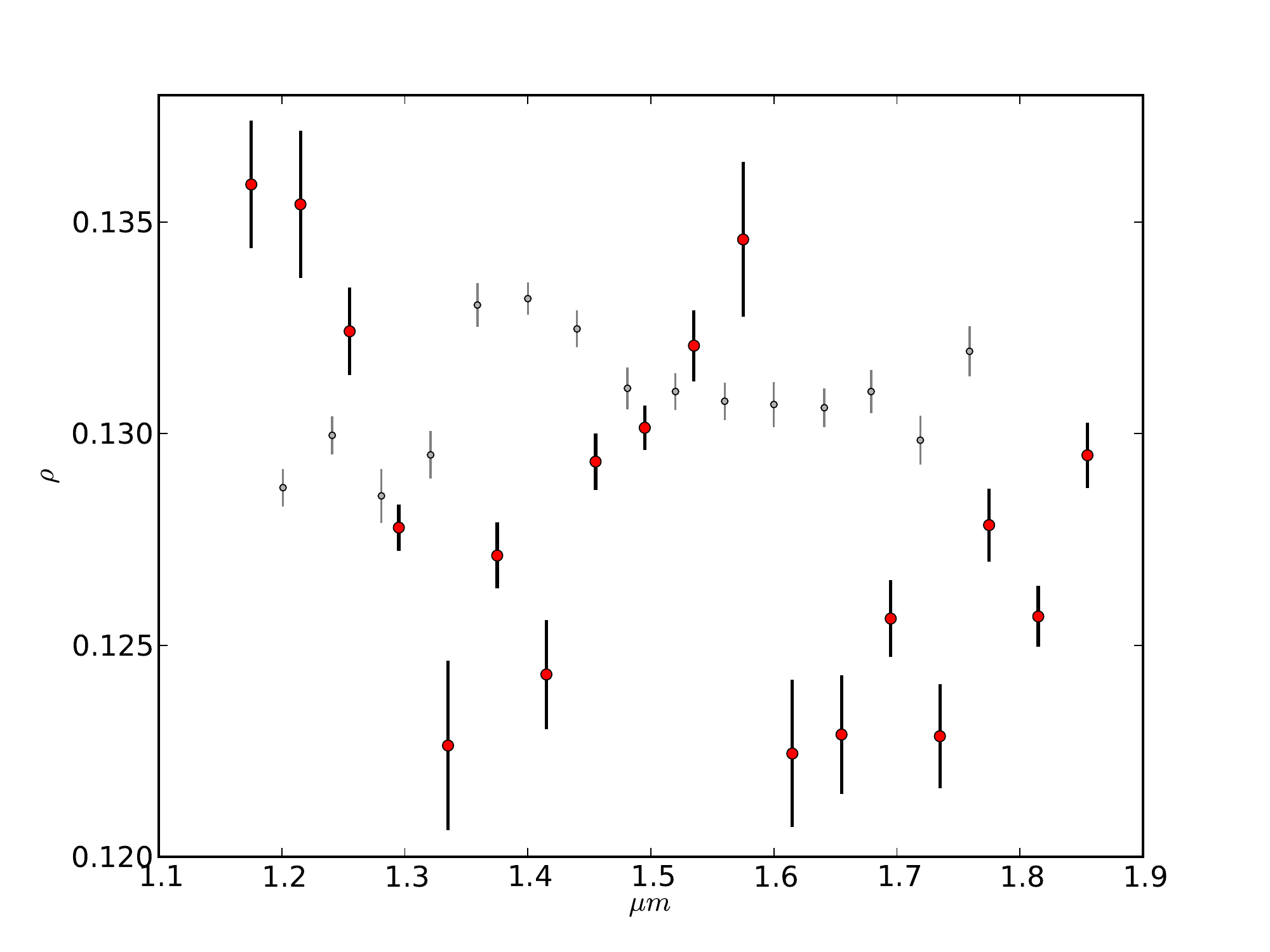}
\caption{Transmission spectrum of XO-1 generated by determining the planet radius from the decorrelated light curves $\rho$ for the 18 wavelengths channels. Out results are the red points, and those from T10 are shown in grey for comparison. The variation is not caused by the atmosphere of the planet, but rather the systematics in the data and the decorrelation method used to correct for these.}
\label{fig:XO1_trans_spec_all}
\end{figure}

Fig.~\ref{fig:XO1_trans_spec_all} shows the resulting transmission spectrum for XO-1. The results from T09 are also plotted for comparison. The amplitude in terms of flux is $>$0.3 \%, and the corresponding uncertainties are about 0.05\%. Both of these are significantly larger than those reported in T10, and obviously we have not reached the precision required to detect any molecular species. The reason for this is clearly related to the decorrelation parameters. Unfortunately, T10 did not provide a plot of their decorrelation parameters, or plots of the light curves. This makes it difficult to check where our reductions differ. As T10 did not mention the need to extrapolate for any of the decorrelation parameters, we must presume the difference between our analysis and theirs resides chiefly in the extraction of the decorrelation parameters.

However, recently \citet{Burke_2010} reported a broadband analysis of this light curve, and provided plots of $\Delta X$, $\Delta Y$ and $\theta$. The slight differences in $\Delta X$ and $\Delta Y$ are probably related to the extraction technique as discussed in Sect.~\ref{sect:HD189733_observations}. The values reported for $\theta$ agree with those in this paper. This confirms the need to extrapolate for the in-transit orbits, in particular for $\theta$, which earlier we concluded was the most important decorrelation parameter for HD 189733.

\section{Summary and discussion}
\label{sect:summary}

We report a re-analysis of transmission spectra observed with NICMOS for HD 189733, GJ-436 and XO-1. We used a linear decorrelation model, similar to that used in other HST analyses \citep[e.g.][S08, P09, T10]{Pont_2008}, where possible, to decorrelate the light curves as they all suffer from considerable systematics. We used a residual permutation algorithm in an effort to obtain more realistic uncertainties than in the literature, although this does not account for uncertainties which arise from orbit-to-orbit flux offsets.

For HD 189733, we find a similar transmission spectrum to S08 when following their procedure as closely as possible. However, using residual permutation, the uncertainties we calculate are often significantly larger, thus reducing the significance of the spectral features. We show that significant systematic noise remains in the in-transit orbits, which is not fully removed by the linear decorrelation model, or the channel-to-channel correction. The shape and amplitude of the spectral features are also strongly dependent on the choice of orbits, decorrelation parameters, and the nature of the model assumed for the baseline function. Given there is no physical reason to assume the baseline flux should follow a linear function of a particular set of optical state parameters, there is no reason to prefer one model over another. We take this as evidence that the linear decorrelation model used is not a robust method to remove systematic effects from the light curves, and that the baseline function in the case of NICMOS grism data cannot reliably be given by a linear function of the optical state vectors from S08.

\begin{figure*}
\includegraphics[width=168mm]{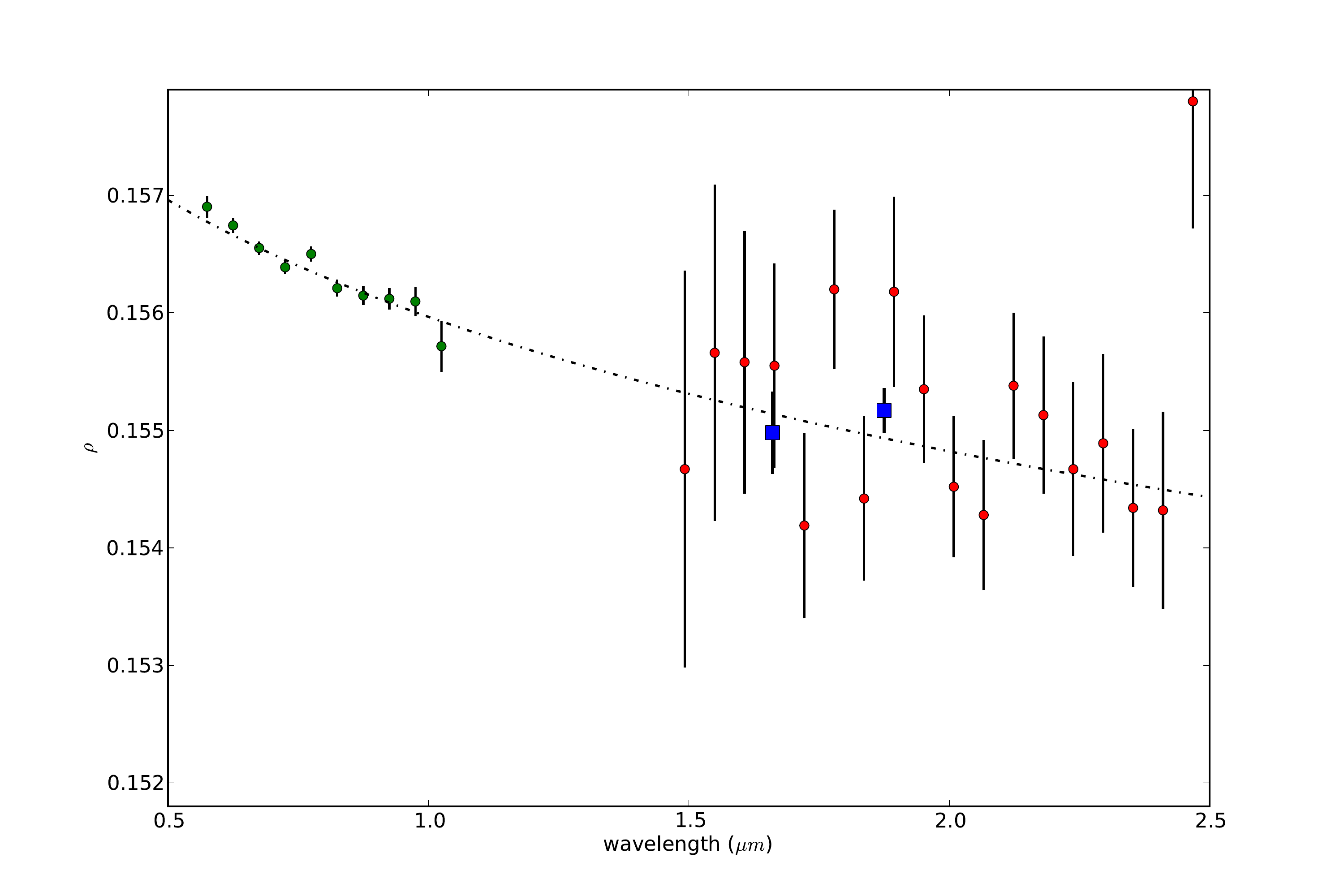}
\caption{Overall transmission spectrum of HD 189733, including the ACS optical data from \citet[][green circles]{Pont_2008}, the NICMOS photometric measurements from \citet[][blue squares]{Sing_2009}, and a re-reduction of G206 NICMOS grism data using only orbits 2 and 4 from this paper (red circles). The dotted-dashed line is the haze given in \citet{Lecavelier_2008}, which the \citet{Pont_2008} and \citet{Sing_2009} results are consistent with. The NICMOS grism data is consistent with the haze using \emph{this} reduction. We emphasise that this case is in no way special, rather we want to illustrate that the NICMOS transmission spectroscopy data may still be consistent with the haze interpretation rather than the reported molecular species, given the difficulty in propagating the effects of systemics to the final uncertainties.}
\label{fig:HD189733_overall_trans_spec}
\end{figure*}

Fig.~\ref{fig:HD189733_overall_trans_spec} shows a plot of the overall transmission spectrum of HD 189733, including the ACS optical data from \citet{Pont_2008}, the NICMOS photometric measurements from \citet{Sing_2009}, and the reduction using only orbits 2 and 4 from this paper. This shows that we can interpret the NICMOS transmission spectrum as consistent with the optical haze, if we adopt this reduction. We emphasise that our analysis does not rule out the presence of molecules in the NIR transmission spectrum of HD 189733, but merely shows that the detection of molecular species is highly dependent on the decorrelation method used.

For GJ-436 we find an amplitude of about 0.07 \% absorption in the transmission spectrum, which is similar to that seen in P09; however, the overall levels are inconsistent. We also find the transmission spectrum is inconsistent with the transit depth of the white light curve. The latter is much more robust given that the integrated white light curve suffers from fewer systemics. The transmission spectrum is consistent with the white light curve when only decorrelating using the orbital phase parameters. This suggests that decorrelating with more parameters, particularly those with offsets in the parameter values, induces spurious offsets in the transmission spectrum. Generally, we conclude that it is unstable to decorrelate HST data using decorrelation parameters that contain offsets between each orbit, as these offsets in easily be transferred to spurious offsets in flux levels of the in-transit orbits. The fact that we can produce a variable but unphysical transmission spectrum with the linear decorrelation model, suggests that we cannot extract useful spectral information below about $0.1\%$ using NICMOS transmission spectroscopy, at least using current methods.

For XO-1 we are unable to reproduce the results from T10, because the decorrelation parameters we measure require an extrapolation for the in-transit orbits to determine the baseline function, and introduce large offsets in the flux levels for different orbits. The best transmission spectrum that we can produce show uncertainties much larger than those reported in T10. We do not have enough information to identify the source of the discrepancy, but we expect that their uncertainties are underestimated, in a similar way to HD 189733. This is all the more puzzling given that \citet{Burke_2010} also produce decorrelation parameters that would require extrapolation for the in-transit orbits. We note the importance of reporting in detail the methods used to correct systematics when they dominate the error budget.

Our re-analysis of all three datasets shows that the systematics in NICMOS transmission spectra cannot reliably be corrected for - at the level needed to detect molecular absorption in hot Jupiters - using the multi-linear decorrelation techniques described in S08 and repeatedly used since. Indeed, there is no physical reason why the systematics should be described by a linear combination of the selected state parameters. In the absence of a better systematics removal technique, there appears to be a $\sim$0.1\,\% floor below which real absorption variation cannot be distinguished from systematics with NICMOS. All claimed detections lie near this level, regardless of object (Fig.~\ref{fig:NICMOS_data}). Whilst this could be coincidental, more evidence is required to support these results, particularly when \citet{Pont_2008} and \citet{Sing_2009} suggest otherwise for HD 189733.

\section*{Acknowledgments}

All of the data presented in this paper were obtained from the Multimission Archive at the Space Telescope Science Institute (MAST). STScI is operated by the Association of Universities for Research in Astronomy, Inc., under NASA contract NAS5-26555. Support for MAST for non-HST data is provided by the NASA Office of Space Science via grant NNX09AF08G and by other grants and contracts. N. P. G and S. A. acknowledge support from STFC grant ST/G002266/2. We are very grateful to D. Sing for providing limb darkening co-efficients for the NICMOS wavelength channels. Finally, we thank the referee, whose insightful comments helped improve the clarity of this paper.


\label{lastpage}

\end{document}